\documentclass[oribibl,runningheads]{llncs}

\usepackage{latexsym}
\usepackage{amsmath}
\usepackage{amssymb}
\usepackage{epsfig}

\newcommand{\ignore}[1]{}

\ignore{
 \setlength{\textwidth}{17cm}
\setlength{\oddsidemargin}{0cm} \setlength{\oddsidemargin}{0cm}
\setlength{\textheight}{20cm} \hoffset -0.2cm \voffset -1cm
 }

\newcommand{\topsepspace}{\vspace{\topsep}}
\long\def\symbolfootnote[#1]#2{\begingroup%
\def\thefootnote{\fnsymbol{footnote}}\footnote[#1]{#2}\endgroup}

\newcommand{\boxtheorem}{\hfill\ensuremath{\Box}}

\newcommand{\nit}[1]{{\it #1}}

\newcommand{\IC}{\nit{IC}}

\newcommand{\defproof}[2]{{\noindent\textbf{Proof of #1:\
}}#2 \boxtheorem \vspace{2mm}\\ \noindent
\ignore{\newline\topsepspace}}

\newcommand{\dproof}[2]{{\noindent\textbf{Proof:\
}}#2 \boxtheorem \ignore{\newline\topsepspace}}

\title{\vspace*{-0.8cm}
Complexity of Consistent Query Answering in Databases under
Cardinality-Based and Incremental Repair Semantics\\
(extended version)
}
\author{{\bf Andrei Lopatenko}~~~~~~~~~~~~~~~~~~{\bf Leopoldo
Bertossi}\\
\hspace*{-9mm}Free University of Bozen-Bolzano~~~~~~~~~~Carleton University\\
Faculty of Computer Science ~~~~~~~~~~~School of Computer
Science\\
\hspace*{-3mm}Bozen-Bolzano, Italy. ~~~~~~~~~~~~~~~~~~~Ottawa, Canada.\\
\hspace*{5mm}\texttt{lopatenko@inf.unibz.it}
~~~~~~~~~~~\texttt{bertossi@scs.carleton.ca}}

\institute{}
\titlerunning{Complexity of Consistent Query Answering}
\authorrunning{Lopatenko and Bertossi}
\begin{document}
 \maketitle


\vspace{-6mm}\begin{abstract} A database $D$ may be inconsistent
wrt a given set $\IC$ of integrity constraints.
  Consistent Query Answering (CQA) is the problem of computing
  from $D$ the answers to a query  that are consistent wrt $\IC$.
  Consistent answers  are invariant under  all the {\em repairs} of $D$, i.e.
 the consistent instances that minimally depart from $D$.
Three classes of repair have been considered in the literature:
those that minimize set-theoretically the set of tuples in the
symmetric difference; those that minimize the changes of attribute
values, and those that minimize the cardinality of the set of
tuples in the symmetric difference. The latter class has not  been
systematically investigated. In this paper we obtain algorithmic
and complexity theoretic results for CQA under this
cardinality-based repair semantics. We do this in the usual,
static setting, but also in a  dynamic framework where a
consistent database is affected by a sequence of updates, which
may make it inconsistent. We also establish comparative results
with the other two kinds of
  repairs in the dynamic case.
\end{abstract}

\section{Introduction}

The purpose of {\em consistent query answering} (CQA) is to
compute query answers  that are consistent with certain integrity
constraints (ICs) that the database as a whole may fail to
satisfy. Consistent answers have been characterized as those that
are invariant under minimal forms of restoration of the
consistency of the database \cite{abc-1999,bc-2003}. A particular
and first notion of minimal restoration of consistency was
captured in \cite{abc-1999} in terms of database {\em repairs},
i.e. consistent  database instances that share the schema with the
original database, but differ from the latter by a {\em minimal
set of whole tuples under set inclusion}. In this paper we call
this semantics ``the S-repair semantics", for being set oriented.
In \cite{bc-2003,fm-2005,abc-1999,clr-2003,abchrs-2003,cm-2005},
complexity bounds for CQA under the S-repair semantics have been
reported.

Two other repair semantics naturally arise and have been
considered in the literature. The {\em A-repair semantics} is
based on changing in a minimal way {\em attribute} values in
database tuples in order to restore consistency. CQA under the
A-repair semantics  has also been investigated
\cite{w-2003,franconi,bbfl-2005,flesca}. The  {\em C-repair
semantics} is based on
 repairs of the original database that minimize the
{\em cardinality} of the set of tuples by which the instances
differ \cite{tplp}. This semantics has received much less
attention so far.

\begin{example}\label{ex:reps}
Consider a database schema $P(X,Y,Z)$ with the functional
dependency $X \rightarrow Y$. The inconsistent instance $D =
\{P(a,b,c), P(a,c,d), P(a,c,e)\}$, seen as a set of ground atoms,
has two S-repairs,  $D_1=\{P(a,b,c)\}$ and $D_2=\{P(a,c,d),$
$P(a,c,e)\}$, because the symmetric set differences with $D$,
$\Delta(D,D_1)$ and $\Delta(D,D_2)$, are minimal under set
inclusion. However, only for $D_2$ the cardinality
$|\Delta(D,D_2)|$ of the symmetric set difference is minimum; and
$D_2$ is the only C-repair.

The query $P(x,y,z)$ has consistent answers $(a,c,d)$ and
$(a,c,e)$ under the C-repair semantics (they are classic answers
in the only C-repair), but none under the S- repair semantics (the
two S-repairs share no classic answers). \boxtheorem
\end{example}
The consistent query answers under C-repairs form a superset of
the consistent answers under  S-repairs, because every C-repair is
also an S-repair.  Actually, in situations where the S-repair
semantics  does not give any consistent answers, the C-repair
semantics may return answers. These answers could be further
filtered out according to other criteria at a post-processing
stage. For example, in the extreme case where there is only one
database tuple in semantic conflict with a possibly large  set of
other tuples, the existence of an S-repair containing the only
conflicting tuple would easily lead to an empty set of consistent
answers. The C-repair semantics would not allow such a repair
(c.f. Example \ref{ex:comp} below).

Furthermore, the C-repair semantics has the interesting property
that CQA, a form of {\em cautious} or \emph{certain} reasoning
(declaring true what is true in {\em all} repairs), and its {\em
brave} or \emph{possible} version (i.e. true in {\em some}
repair), are mutually reducible in polynomial time and share the
same data complexity. This is established in Section
\ref{sec:mcss} by proving first some useful graph-theoretic lemmas
about maximum independent sets that are interesting in themselves,
and have a wider applicability in the context of CQA.

 In \cite{tplp},  C-repairs were specified using disjunctive logic programs with
stable model semantics \cite{gelLif91} and weak cardinality
constraints \cite{buccafurri2k}. In this paper, applying the
graph-theoretic techniques and results mentioned above,
 we
obtain the first non-trivial complexity results for CQA under the
C-repair semantics. Our emphasis is on CQA, as opposed to
computing or checking specific repairs.

All the complexity bounds on CQA given  so far in the literature,
no matter which repair semantics is chosen, consider  {\em the
static case}: Given a snapshot of a database, a set of integrity
constraints, and a query, the problems are the computation and
verification of consistent answers to the query. In this paper we
also take into account dynamic aspects of data, studying the
complexity of CQA when the consistency of a  database may be
affected by update actions.

\begin{example}\label{ex:inc} (example \ref{ex:reps} continued)
The C-repair $D_2 = \{P(a,c,d),$ $P(a,c,e)\}$ is obviously
consistent, however after the execution of the update operation
$\nit{insert}(P(a,f,d))$ it becomes inconsistent. In this case,
the only C-repair of $D_2 \cup \{P(a,f,d)\}$ is $D_2$ itself. So,
CQA from $D_2 \cup \{P(a,f,d)\}$ amounts to classic query
answering from $D_2$. However, if we start from the consistent
instance $D' = \{P(a,c,d)\}$, executing the same update operation
leads to two C-repairs, $D'$ and also $\{P(a,f,d)\}$, and now CQA
from $D' \cup \{P(a,f,d)\}$ is different from classic query
answering from $D'$, because two repairs have to be considered.
\boxtheorem
\end{example}
Understanding and handling CQA in a dynamic setting is crucial for
its applicability. Incremental methods should be developed, since
it would be inefficient to compute a materialized repair of the
database or a consistent answer to a query from scratch after
every update.

While we think that the right repair semantics may be application
dependent, being able to compare the possible semantics in terms
of complexity may also shed some light on what may be the repair
semantics of choice. This comparison should consider both static
and incremental CQA, because a specific semantics might be better
than others in terms of complexity when the database is affected
by certain updates. In this paper we compare the C-repair
semantics with the S- and A-repair semantics mentioned before,
 and both in the static and incremental
settings.

In Section \ref{sec:mcss} we prove that static CQA under C-repairs
is $P^\nit{NP(log(n))}$-hard for denial constraints and ground
atomic queries; which contrasts with the $\nit{PTIME}$ result for
S-repairs in \cite{cm-2005}. On the other side, in Section
\ref{sec:incr}, we prove that incremental CQA, i.e. CQA in the
dynamic setting, under the C-repair semantics is in  $\nit{PTIME}$
for denial constraints and conjunctive queries; and that the same
problem under S-repairs is $\nit{coNP}$-hard (in data).

The naive algorithms for incremental CQA under the C-repair
semantics are polynomial in data, but exponential in the size of
the update sequence. In consequence, we also study the  {\em
parameterized complexity} \cite{downfel,flum} of incremental CQA
under
 the C-repair semantics, being the parameter the size of
the update sequence. We establish that the problem is {\em fixed
parameter tractable} (FPT).

For establishing comparisons with the C-repair semantics, we
obtain new results on the static and incremental complexity  both
under the classic, i.e. S-repair semantics, and the A-repair
semantics. We prove, for the former, that incremental CQA is
coNP-hard; whereas for the latter, static and incremental CQA
become both $P^\nit{NP}$-hard in data.

We concentrate on relational databases and denial integrity
constraints, which include most of the constraints found in
applications where inconsistencies naturally arise, e.g.
census-like databases \cite{bbfl-2005}, experimental samples
databases, biological databases, etc.

Complexity results in this work refer all to
data complexity. For complexity theory we refer to
\cite{papadimitriou94}; and to \cite{flum} for parameterized
complexity. 
However,
we briefly recall some of the complexity classes used in this
paper. $\nit{FP}$ is the class of functional problems  that are
solvable in polynomial time. $P^\nit{NP}$  (or $\Delta_2^P$) is
the class of decision problems solvable in polynomial time by a
machine that makes calls to an $\nit{NP}$ oracle.
$P^\nit{NP(log(n))}$ is similarly defined, but the number of calls
is logarithmic. It is not known if $P^\nit{NP(log(n))}$ is
strictly contained in $P^\nit{NP}$. The functional class
$\nit{FP}^\nit{NP(log(n))}$ is similarly defined. The class
$\Delta^P_3\!(\nit{log(n)})$ contains decision problems that can
be solved by a polynomial time machine that makes a logarithmic
number of calls to an oracle in $\Sigma^P_2$. For more details we
refer to \cite{papadimitriou94}; and to \cite{flum} for
parameterized complexity.

\section{Semantics for Consistent Query Answering}

A relational database instance $D$ is a finite set of ground atoms
$R(\bar{t})$ (also called {\em database tuples}\footnote{We also
use the term {\em tuple} to refer to a finite sequence $\bar{t} =
(c_1, \ldots, c_n)$ of constants of the database domain $\cal U$,
but a {\em database tuple} is a ground atomic sentence with
predicate in $\cal D$ (excluding built-ins predicates, like
comparisons).}), where $R$ is a relation in the schema ${\cal D}$,
and $\bar{t}$ is a finite sequence of constants from the
 domain $\cal U$.  A database atom is of the form
$R(\bar{t})$, where $R$ is a predicate in $\cal D$, and $\bar{t}$
may contain constants or variables. A database literal is a
database atom or a negation of a database atom. With
$\Delta(D',D)$ we denote the symmetric difference
$(D'\smallsetminus D) \cup (D\smallsetminus D')$ between instances
$D, D'$, conceived both as sets of ground atoms.

The relational schema ${\cal D}$
determines a first-order language $L({\cal D})$ based on the
relation names, the elements of $\cal U$, and extra built-in
predicates.
 In the language $L({\cal D})$, integrity constraints are sentences, and queries are
 formulas, usually with free variables. We assume in this paper
 that sets $\IC$ of ICs are always consistent in the sense that
 they are simultaneously satisfiable as first-order
 sentences.
A database is {\em consistent} wrt to a given set of integrity
constraints $\IC$ if the sentences in $\IC$ are all true in $D$,
denoted $D \models \IC$. An answer to a query $Q(\bar{x})$, with
free variables $\bar{x}$, is a tuple  $\bar{t}$ that makes $Q$
true in $D$ when the variables in $\bar{x}$ are interpreted as the
corresponding values in $\bar{t}$, denoted $D \models Q[\bar{t}]$.

\begin{definition}\label{def:rep}  \em For a database $D$,
  integrity constraints  $\IC$, and a partial order $\preceq_{D,{\cal S}}$ over
  databases that depends on the original database $D$ and a repair
  semantics ${\cal S}$, a \emph{repair of $D$ wrt $\IC$ under ${\cal S}$} is an
  instance $D^\prime$ such that: (a) $D^\prime$ has the same schema
  and domain as $D$; (b) $D^\prime \models \IC$; and (c) there is no $D''$ satisfying (a) and (b), such
  that $D'' \prec_{D,{\cal S}} D^\prime$, i.e. $D'' \preceq_{D,{\cal S}} D^\prime$ and not
  $D^\prime \preceq_{D,{\cal S}} D''$.
  The set of all repairs is denoted with $\nit{Rep}(D, \IC,{\cal S})$. \boxtheorem
\end{definition}
The class $\nit{Rep}(D, \IC,{\cal S})$ depends upon  the semantics
$\cal S$, that determines the partial order  $\preceq$
 and the way repairs can be
obtained, e.g. by allowing both insertions and deletions of whole
database tuples \cite{abc-1999}, or deletions of them only
\cite{cm-2005}, or only changes of attribute values
\cite{w-2003,bbfl-2005,flesca}, etc. (c.f. Definition
\ref{def:dist}.)
 We summarize  here the  most common repair semantics.
\begin{definition} \label{def:dist}  \em
(a) \emph{S-repair semantics} \cite{abc-1999}: $D' \preceq_{D,S}
D''$
iff $\Delta(D',D) \subseteq \Delta(D'',D)$.\\
    (b) \emph{C-repair semantics}: $D' \preceq_{D,C} D''$ iff $|\Delta(D',D)| \leq
    |\Delta(D'',D)|$.\\
    (c) \emph{A-repair semantics}: $D' \preceq_{D,A} D''$ iff $f(D,D') \leq
f(D,D'')$, where $f$ is a fixed numerical aggregation function
over differences of attribute values. \boxtheorem
\end{definition}
More details about the A-repair semantics can be found in Section
\ref{sec:attrib}. Particular cases of A-repairs can be found in
\cite{franconi,flesca}, where the aggregation function to be
minimized is the number of all attribute changes; and in
\cite{bbfl-2005}, where the function is the overall quadratic
difference obtained from the changes in numerical attributes
between the original database and the repair. S-repairs and
C-repairs are ``tuple-based", in the sense that consistency is
restored by inserting and/or deleting whole database tuples;
whereas A-repairs are obtained by changing attributes values in
existing tuples only.

In Example \ref{ex:reps},
 attribute-based repairs could be $\{P(a,c,c), P(a,c,d),$ $P(a,c,e)\}$,
 suggesting that we made a mistake in the second argument of the first tuple, but
also $\{P(a,b,c), P(a,b,d),$ $ P(a,b,e)\}$. If the aggregate
function in Definition \ref{def:dist}(c) is the number of changes
in attribute values, the former would be a repair, but not the
latter. A-repairs may not be S- or C-repairs if the changes of
attribute values have to be simulated via deletions followed by
insertions.

\begin{definition}
\label{def:semantics} \em Let $D$ be a database, $\IC$
  a set of ICs, and $Q(\bar{x})$ a query.  (a) A ground tuple
  $\bar{t}$ is a \emph{consistent answer} to $Q$ wrt $\IC$ under semantics $\cal S$
  if for every $D' \in \nit{Rep}(D,\IC,\cal S)$,
  $D' \models Q[\bar{t}]$.~  (b)
  $\nit{Cqa}(Q,D,\IC,{\cal S})$ is the set of consistent answers to
  $Q$ in $D$ wrt $\IC$ under semantics ${\cal S}$. If $Q$ is a sentence (a boolean query),
  $\nit{Cqa}(Q,D,\IC,{\cal S}) := \{\nit{yes}\}$ when $D' \models Q$ for
  every $D' \in \nit{Rep}(D,\IC,\cal S)$, and
  $\nit{Cqa}(Q,D,\IC,{\cal S}) := \{\nit{no}\}$, otherwise. (c)
  $\nit{CQA}(Q,\IC,{\cal S}) := \{(D,\bar{t}) ~|~ \bar{t} \in
  \nit{Cqa}(Q,D,\IC,{\cal S})\}$ is the \emph{decision problem of consistent
    query answering}.
\boxtheorem
\end{definition}
Denial constraints are integrity constraints expressed by $L({\cal
D})$-sentences
 of the form
    $\forall \bar{x} \neg(A_1\land\ldots\land A_m \wedge \gamma),$
where each $A_i$ is a database atom and $\gamma$ is a conjunction
of comparison atoms.
In particular, functional dependencies (FDs), e.g. $\forall x \forall y
\forall z
\neg(R(x,y) \wedge R(x,z) \wedge y \neq z)$, are denial
constraints.  For denial ICs,
tuple-based repairs are obtained by tuple deletions only
\cite{cm-2005}.

\section{Complexity of CQA under the C-Repair Semantics} \label{sec:mcss}

As a consequence of the specification of C-repairs as the stable
models of disjunctive logic programs with non-prioritized weak
constraints \cite{tplp} and the results in \cite{buccafurri2k}, we
obtain that an upper bound on the data complexity of CQA under the
C-repair semantics is the class $\Delta^P_3\!(\nit{log(n)})$.

In \cite{abchrs-2003}, {\em conflict graphs} were first introduced
 to study the
  complexity of CQA for aggregate queries wrt FDs under the S-repair semantics. They
 have as vertices the database tuples; and edges connect
 two tuples that simultaneously violate a FD. There is a one-to-one correspondence between
 S-repairs of the database and the set-theoretically
 {\em maximal} independent sets  in the conflict graph.
 Similarly,  there is  a one-to-one correspondence
 between C-repairs and {\em maximum} independent sets  in the same graph (but now
 they are maximum in cardinality).

 Conflict graphs for databases wrt general denial
constraints become {\em conflict hypergraphs} \cite{cm-2005} that
have as vertices the database tuples, and as hyperedges the (set
theoretically minimal) collections of tuples that simultaneously
violate one of the denial constraints. The size of the hypergraph
(including vertices and hyperedges) is polynomial in the size of
the database, because we have a fixed set of denial constraints.
The correspondence for conflict graphs between repairs and
independent sets $-$maximum or maximal depending on the
semantics$-$ still holds for hypergraphs, where an independent set
in an hypergraph is a set of vertices that does not contain any
hyperedges \cite{cm-2005}.

 Notice that, unless an IC forces a particular tuple
 not to belong to the database,\footnote{We do not consider in this work such {\em non generic} ICs
 \cite{bc-2003}.} every tuple in the original database belongs to
 some
 S-repair, but not necessarily to a C-repair (c.f. Example \ref{ex:reps}, where the
 tuple $P(a,b,c)$ does not belong to the only C-repair).

 In consequence, testing membership of vertices to some maximum
 independent set becomes a relevant for C-repairs.
 The complexity of this problem will determine the
 complexity of CQA under the C-repair semantics.
 For  this purpose we will use some graph-theoretic constructions and lemmas about
 maximum independent sets, whose proofs use a
self-reducibility property of independent sets that can be
expressed as follows: For any graph $G$ and vertex $v$, every
maximum independent set that contains $v$ (meaning maximum among
the independent sets that contain $v$) consists of vertex $v$
together with a maximum independent set of the graph $G'$ that is
obtained from $G$ by deleting all vertices  adjacent to $v$.

To keep the presentation simpler, we concentrate mostly on
conflicts graphs and FDs. However, the results obtained carry over
to denial constraints and their hypergraphs. Notice, as a
motivation for the next lemmas, that a ground atomic query is
consistently true when it belongs, as a database tuple, i.e. as a
vertex in the conflict graph, to all the maximum independent sets
of the conflict graph.

\begin{lemma}\label{lem:graphExt}\em
Consider a graph $G$ and a vertex $v$ in it. (a) For the graph
$G'$ obtained by adding a new vertex $v'$ that is connected only
to the neighbors of $v$, the following properties are equivalent:
1. There is a maximum independent set of $G$ containing $v$. 2.
$v$ belongs to every maximum independent set of $G'$. 3. The
sizes of maximum independent sets in $G$ and $G'$ differ by one.\\
(b) There is a graph $G'$ extending $G$ that can
 be  constructed in logarithmic space, such that
 $v$ belongs to all maximum independent sets of $G$ iff $v$
 belongs to some maximum independent set of $G'$.
\end{lemma}

\dproof{Lemma \ref{lem:graphExt}}{(a) We consider the three cases for membership
of $v$ to maximum
independent sets in $G$. Let $m$ be the cardinality of a maximum
independent set in $G$. We establish now the first bi-conditional.
The second
bi-conditional follows directly from the analysis for the first one. \\
(a) Assume that $v$ belongs to a maximum independent set $I$ of
$G$. In this case, $v'$ can be added to $I$ obtaining an
independent set of $G'$. In this case $|I \cup \{v'\}| \geq m+1$.

Assume that $v$ does not belong a some maximum independent set
$I'$ of $G'$. If $v \notin I'$, then some of its neighbors belong
to $I'$, and then, $v' \notin I'$. In consequence, $I'$ is also a
maximum independent set of $G$. Then, $|I'| = m$. But this is not
possible, because the size of independent set of $I'$ is at least
$m +1$.\\
(b) Assume that $v$ does not belong to any maximum independent
sets of $G$. Then, some of it neighbors can be found in every
maximum independent set of $G$, and none of them can be extended
 with $v'$ to become an independent set of $G'$.

 So, all the maximum independent set of $G$ are maximum independent sets of
 $G'$ of size $m$.

 Assume, that $v$ belongs to all maximum independent sets of $G'$.
  Then none of the neighbors of $v$ can be found in
 independent sets of $G$, and then
$v'$ can be found in all the maximum independent sets of
 $G'$. Since the maximum independent sets of $G'$ have at least
 cardinality $m$, it must hold that the maximum independent sets
 of $G'$ have cardinality at least $m+1$. Then the deleting $v'$
 from all the maximum independent sets of $G'$ will give us
 independent sets of $G$ of size at least $m$, i.e. maximum
 independent sets of $G$. To all of them $v$ belongs. A
 contradiction.\\
 (b) (sketch) Hang a rhombus from $v$, i.e.
 add three other vertices, two of them connected to $v$, and the
 third one, connected to the two previous ones. Then,
 reason by cases as in the proof of part (a).}

From this lemma and the membership to $\nit{FP}^\nit{NP(log(n))}$
of computing the size of a maximum clique in a graph
\cite{krentel}, we obtain

 \begin{lemma}\label{lem:dec-graphNPlogn} \em The problems of deciding
  for a vertex in a graph if it belongs to some maximum
  independent set and
  if it belongs to all maximum
  independent sets are both in $P^\nit{NP(log(n))}$. 
\end{lemma}

\dproof{Lemma \ref{lem:dec-graphNPlogn}}{For the first claim, given a
graph $G$ and a vertex $v$, build in polynomial time the graph
$G'$ as in Lemma \ref{lem:graphExt}(a). It holds that $v$ belongs
to some maximum independent set of $G$ iff $v$ belongs to every
maximum independent set of $G'$. Now, $v$ belongs to every maximum
independent set of $G'$ iff $|\mbox{maximum independent set}$
$\mbox{in } G'| - |\mbox{maximum independent set in } G| = 1$.

Since computing the maximum cardinality of a clique can be done in
time $\nit{FP}^\nit{NP(log(n))}$ \cite{krentel} (see also
\cite[theorem 17.6]{papadimitriou94}), computing the maximum
cardinality of an independent set can be done in the same time
(just consider the complement graph). In consequence, in order to
decide about $v$ and $G$, we can compute the cardinalities of the
maximum independent set for $G$ and $G'$ in 2 times
$\nit{FP}^\nit{NP(log(n))}$, and next compute their difference. It
total, we can perform the whole computation in
$\nit{FP}^\nit{NP(log(n))}$. In consequence, by definition of
class $\nit{FP}^\nit{NP(log(n))}$, we can decide by means of a
polynomial time machine that makes $O(\nit{log}(n))$ calls to an
$\nit{NP}$ oracle, i.e. the decision is made in time
$P^\nit{NP(log(n))}$. The same proof works for the second claim.
It can also be obtained from the first claim and Lemma
\ref{lem:graphExt}(b).}

\begin{theorem}\label{prop:memb}
\em For functional dependencies and ground atomic queries,  CQA
under the C-repair semantics belongs to $P^\nit{NP(log(n))}$.
\end{theorem}

\dproof{Theorem \ref{prop:memb}}{Construct the conflict
graph for the instance wrt the FDs. An atomic ground query is
consistently true if the corresponding vertex in the conflict
graph belongs to all the maximum independent sets.  Then use
Lemma \ref{lem:dec-graphNPlogn}.}\\

Considering the maximum independent sets, i.e. C-repairs, as a
collection of possible worlds,
 the previous lemma shows a close connection between the {\em
 certain} C-repair semantics (true in {\em every} repair), that is
 the basis for CQA, and the
 {\em possible} C-repair semantics (true in {\em some} repair).
CQA under these semantics and functional dependencies are
polynomially reducible
 to each other; actually also for negations of ground atomic queries.

 \begin{lemma} \label{lemma:allRed} \em
 The following problems are mutually
 $\nit{LOGSPACE}$-reducible to each other:~
 (1) {\em Certain positive:}~ Given a vertex $v$ and a graph $G$, decide if
$v$ belongs to every maximum independent set of $G$.~ (2) {\em
Certain negative:}~ Given a vertex $v$ and a graph $G$, decide if
all the maximum independent sets of $G$ do not contain $v$.~ (3)
{\em Possible negative:}~ Given a vertex $v$ and a graph $G$,
decide if there is a maximum independent set of $G$ that does not
contain $v$.~ (4) {\em Possible positive:}~ Given a vertex $v$ and
a graph $G$, decide if $v$ belongs to at least one maximum
independent set of $G$. 
\end{lemma}

\dproof{Lemma \ref{lemma:allRed}}{We prove: (1) $\Rightarrow$ (2) $\Rightarrow$ (3) $\Rightarrow$
(4). That (4) $\Rightarrow$ (1) was established in Lemma
\ref{lem:graphExt}(a).\\
(1) $\Rightarrow$ (2):~ Given a graph $G$ and a vertex $v$, extend
$G$ to  a graph $G'$ by adding  new vertices $s, s'$ with $s$
connected to $v$ and $s'$ connected to $s$.
If $v$ belongs to every MIS of $G$, then $s$ does not belong to any MIS of $G'$.\\
If $v$ belongs to one MIS of $G$, but  not to the others, then $s$ belongs to some MIS of $G'$
 and does not belong any other MIS of $G'$.
If $v$ does not belong to any MIS of $G$, then $s$  belongs to one
MIS of $G'$ but not to the others. Thus, $v$ belong to every MIS
of $G$ if and only if $s$ does not belong to any MIS of $G'$.
\\
(2) $\Rightarrow$ (3):~ Given $G, v$, extend $G$ to $G'$ by adding
a vertex $s$ and connecting it to $v$. If $v$ belongs to every MIS
of $G$, then $s$ belongs to every MIS of $G'$. If $v$ belongs to
one MIS of $G$ but and not to the others, then $s$ belongs to
every MIS of $G'$. If $v$ does not belong to any MIS of $G$, then
either $s$ does not belong to any MIS of $G'$ or there is a MIS of
$G'$  to which $s$ does not belong (depending on $G$).
\\
(3) $\Rightarrow$ (4):~ Given $G, v$, extend $G$ to $G'$ by adding
vertices $s_1, s_2, s_3, s$, and the edges $\{s_1,v\},
\{s_2,s_1\}, \{s_3,s_1\}, \{s,s_2\}, \{s,s_3\}$. If $v$ belongs to
every MIS of $G$, then $s$ does not belong to any MIS of $G'$. If
$v$ belongs to one MIS of $G$ but not to the others, then $s$
belongs to one MIS of $G'$ but not  to the others. If $v$ does not
belong to any MIS of $G$, then $s$ belongs to one MIS of $G'$ but
not to the others. }\\

Since the negation $\neg R(\bar{t})$ of a ground atomic query
$R(\bar{t})$ is consistently true wrt the C-repair semantics iff
the vertex corresponding to $R(\bar{t})$ in the conflict graph does not
belong to any maximum independent set, using Lemma
\ref{lemma:allRed} we can extend
 Theorem \ref{prop:memb} to conjunctions of literals.\footnote{This can also be obtained, less
 directly, from the closure of $P^\nit{NP(log(n))}$
 under complement.}\ignore{  Notice that a
vertex does not belongs to any maximum independent sets means that
the {\em certain answer} to the corresponding negated query is
{\em yes} and the {\em possible answer} to the corresponding
atomic query is {\em no} (or better, {\em false} in the sense that
it is false in every C-repair). On the other side, that there is a
maximum independent set to which a vertex does not belong means
that the {\em possible answer} to the corresponding negated query
is {\em yes} and the {\em certain answer} to the positive query is
{\em no}. Theorem \ref{prop:memb} also holds for queries that are
conjunctions of atoms.} Actually, since
  Lemmas \ref{lem:graphExt},
\ref{lem:dec-graphNPlogn} and \ref{lemma:allRed} still hold for
hypergraphs,  we obtain

\begin{theorem}\label{prop:membDCs} \em
For denial constraints and queries that are  conjunctions of
literals, CQA under the C-repair semantics belongs to
$P^\nit{NP(log(n))}$. 
\end{theorem}

\dproof{Theorem \ref{prop:membDCs}}{We use the
conflict hypergraph. The problem of determining the maximum clique
size for hypergraphs is in $\nit{FP}^{\nit{NP}(\nit{log}(n)}$ by
the same argument as for conflict graphs: Deciding  if the size of
maximum clique is greater than $k$ is in $\nit{NP}$. So, by asking
a logarithmic number of $\nit{NP}$ queries, we can determine the
size of maximum clique.

The membership to $P^{\nit{NP}(\nit{log}(n))}$ of CQA for the
C-repair semantics
 still holds for conjunctive queries
without existential variables. In fact, given an inconsistent
database $D$, a query $Q$, and a ground tuple $t$, we check if $t$
is consistent answer to $Q$ from $D$ as follows: Check if $t$ is
an ordinary answer to $Q$ in $D$ (without considering the
constraints). If not, the answer is {\em no}.

Otherwise, let $t_1, \ldots, t_k$ be the database tuples which are
answers to $Q$ in $D$ and produce $t$ as an answer. Since $Q$ does
not contain existential variables, only one such set exists.
Compute the size of a maximum independent set for the graph
representation of $D$, say $m_0$. Compute the size of a maximum
independent set for the graph representation of $D \smallsetminus
\{t_1\}$, say $m_1$. If $m_1 = m_0$, then there exist a maximum
independent set of $D$ that does not contain $t_1$. So, there
exists a minimum repair that does not satisfy that $t$ is an
answer to $Q$. If $m_1 < m_0$, repeat this procedure for all
tuples in $t_1, \ldots, t_k$. Thus, we have to pose $k$ queries
(that is determined only by the size of the query) to an
$\nit{FP}^{\nit{NP(log}(n))}$ oracle.

As a consequence,  CQA for conjunctive queries without existential
variables is in $\nit{P}^{\nit{NP(log}(n))}$.}\\

Now we will represent the
 maximum independent sets of a graph as  C-repairs of an
inconsistent database wrt a denial constraint. This is
interesting, because conflict graphs for databases wrt denial
constraints are, as indicate before, actually conflict
hypergraphs.
 \begin{lemma}\label{lem:everyGraph}\em
 There is a fixed database schema ${\cal D}$ and a denial constraint $\varphi$ in
 $L({\cal D})$, such that for every graph $G$, there is an instance
 $D$ over ${\cal D}$, whose C-repairs wrt $\varphi$ are in
 one-to-one correspondence with the maximum independent sets of
 $G$. Furthermore, $D$ can be built in polynomial time in the size
 of $G$.\footnote{We thank Phokion Kolaitis for pointing to an issue in the original proof, and allowing us to highlight
the \nit{coNP}-completeness of cardinality repair checking implicit in this result (c.f. Corollary \ref{cor:checking}).} 
 \end{lemma}


  \dproof{Lemma \ref{lem:everyGraph}}{Consider a graph $G = \langle V,
E\rangle$, and assume the vertices of $G$ are uniquely labelled.
Consider the database schema with three relations, $\nit{Vertex}(v)$, \linebreak $\nit{Edges}(v_1, v_2, e)$, and $N(e)$; and the denial constraint $\forall
v_1 v_2 e \neg(\nit{Vertex}(v_1) \land$ \linebreak $\nit{Vertex}(v_2) \land
\nit{Edges}(v_1, v_2,e) \land N(e))$. $\nit{Vertex}$ stores the vertices of
$G$. For each edge $\{v_1,v_2\}$ in $G$, $\nit{Edges}$ contains
$n$ tuples of the form $(v_1, v_2,i)$, where $n$ is the number of
vertices in $G$. All the values in the third attribute of
$\nit{Edges}$ are different, say from $1$ to $n |E|$. Relation $N$ stores the edges appearing in the the third attribute of $\nit{Edges}$. The size of
the database instance obtained trough this padding of $G$ is still
polynomial in size.

This instance is highly inconsistent, and  its C-repairs are all
obtained by deleting vertices, i.e. elements of $\nit{Vertex}$
alone. In fact, an instance such that all tuples but one in
$\nit{Vertex}$ are deleted, but all tuples in $\nit{Edges}$ are
preserved is a consistent instance. In this case, $n-1$ tuples are
deleted. If we try to achieve a repair by deleting tuples from
$\nit{Edges}$, say $(v_1,v_2,i)$, then in every repair of that
kind all the $n$ tuples of the form $(v_1, v_2,j)$ have to be
deleted as well. This would not be a minimal cardinality repair.

Similarly, no repair can be obtained by
deleting tuples from $N$,
since if one tuple is deleted, then $n$ tuples have to be deleted (same
argument as to why
one  cannot get a repair by deleting tuples from $\nit{Edges}$).

Assume that $I$ is a maximum cardinality independent set of $G$.
The deletion of all tuples $(v)$ from $\nit{Vertex}$, where $v$
does not belong to $I$, is a C-repair. Now, assume that $D$ is a
repair. As we know, only tuples from $\nit{Vertex}$ may be
deleted. Since, in order to satisfy the constraint, no two
vertices in the graph that belong to $D$ are adjacent, the
vertices remaining in $\nit{Vertex}$ form an independent set in
$G$.

In general, the number of deleted tuples is equal to $n-|I|$,
where $I$ is an independent set represented by a repair. So each
minimal cardinality repair corresponds to a maximum independent
set and vice-versa.}

\begin{corollary} \label{cor:checking} \em
There is  a denial constraint for which repair checking under the C-repair semantics is \nit{coNP}-complete
(in data).
\end{corollary}

\dproof{Corollary \ref{cor:checking}}{Consider the following maximum independent set problem:

$P_1$: \ Given a graph $G = \langle V(G), E(G)\rangle$, and a set of vertices $V' \subseteq V(G)$,
decide if $V'$ is a maximum independent set of $G$.

It can be reduced to a database repair  problem for the
database schema $S$ and a set of denial constraints $\nit{IC}$
in Lemma \ref{lem:everyGraph}:

$P_0$: \  Given
a pair $<D,D'>$, decide if $D'$ is a C-repair of $D$ wrt. $\nit{IC}$.

The maximum independent set problem ($P_1$)
is \nit{coNP}-hard by reduction from the decision version of the maximum
independent set, namely:

$P_2$: \
Given a graph $G$ and the number $k$, decide if there exist a maximum
independent set $G'$ of $G$ with $|G'| > k$.

$P_2$ can be reduced to $P_1$: \
Given graph $G$ and $k$,
construct graph $G_1$
as $G$ plus  a set of new vertices $v_1, \ldots, v_k$,
with each $v_i$ connected to all vertices of $G$, and no edges
between $v_i, v_j$. The size of this graph is polynomial
in the $|G|, k$. It holds that
{\em there is no} maximum independent set of $G$ of cardinality greater than  $k$ if and only if
$\{v_1,\ldots,v_k\}$ is the maximum independent set of $G_1$.}\\


From Lemma \ref{lem:everyGraph} and the
$P^\nit{NP(log(n))}$-completeness of determining the size of a
maximum clique \cite{krentel}, we obtain
\begin{theorem} \label{cor:size} \em
Determining the size of a C-repair for denial constraints is
complete for $\nit{FP}^\nit{NP(log(n))}$. 
\end{theorem}

\dproof{Theorem \ref{cor:size}}{This follows from Lemma
\ref{lem:everyGraph}, the fact
that C-repairs correspond to maximum cliques in the complement of
the conflict graph \cite{abchrs-2003}, and the
$P^\nit{NP(log(n))}$-completeness of determining the size of a
maximum clique \cite{krentel}.}\\


\ignore{
\vspace*{3.4cm} \psset{xunit=0.5cm,yunit=0.5cm} \hspace*{10mm}
\begin{pspicture}{0,0}{10,8}[h]
\psellipse[linestyle=dashed, fillstyle=none](1,2)(1,2)
\psellipse[linestyle=dashed, fillstyle=none](1,8)(1,2)
\psellipse[linestyle=dashed, fillstyle=none](5,8)(1,2)
\psellipse[linestyle=dashed, fillstyle=none](5,2)(1,2)
\qdisk(8.5,8){2pt} \qdisk(8.5,2){2pt}
\qdisk(5,8){2pt} \qdisk(5.2,8.4){2pt} \qdisk(4.8,7.5){2pt}
\qdisk(4.8,9.6){2pt} \qdisk(4.7,6.3){2pt} \qdisk(5,7){2pt}
\qdisk(5.5,7,4){2pt} \qdisk(4.5,8.3){2pt}
\qdisk(5,2){2pt} \qdisk(5.2,2.4){2pt} \qdisk(4.8,1.5){2pt}
\qdisk(4.8,3.6){2pt} \qdisk(4.7,0.3){2pt} \qdisk(5,1){2pt}
\qdisk(5.5,1,4){2pt} \qdisk(4.5,2.3){2pt}
\psline{-}(8.5,8)(8.5,2)
\psline(8.5,8)(5,8) \psline(8.5,8)(5.2,8.4)
\psline(8.5,8)(4.8,7.5) \psline(8.5,8)(4.8,9.6)
\psline(8.5,8)(4.7,6.3) \psline(8.5,8)(5,7)
\psline(8.5,8)(5.5,7,4) \psline(8.5,8)(4.5,8.3)

\psline(8.5,2)(5,2) \psline(8.5,2)(5.2,2.4)
\psline(8.5,2)(4.8,1.5) \psline(8.5,2)(4.8,3.6)
\psline(8.5,2)(4.7,0.3) \psline(8.5,2)(5,1)
\psline(8.5,2)(5.5,1,4) \psline(8.5,2)(4.5,2.3)
\psline{-}(5.2,8.4)(1.5,7.6) \psline{-}(5.2,8.4)(0.7,8.4)
\psline{-}(5.2,8.4)(1.1,9.3) \psline{-}(5.2,8.4)(0.8,7.2)
\psline{-}(4.7,6.3)(1.5,7.6) \psline{-}(4.7,6.3)(0.7,8.4)
\psline{-}(4.7,6.3)(1.1,9.3) \psline{-}(4.7,6.3)(0.8,7.2)
\psline{-}(4.8,9.6)(1.5,7.6) \psline{-}(4.8,9.6)(0.7,8.4)
\psline{-}(4.8,9.6)(1.1,9.3) \psline{-}(4.8,9.6)(0.8,7.2)
\psline{-}(5,7.5)(1.5,7.6) \psline{-}(5,7.5)(0.7,8.4)
\psline{-}(5,7.5)(1.1,9.3) \psline{-}(5,7.5)(0.8,7.2)
\psline{-}(5.2,2.4)(1.5,1.6) \psline{-}(5.2,2.4)(0.7,2.4)
\psline{-}(5.2,2.4)(1.1,3.3) \psline{-}(5.2,2.4)(0.8,1.2)
\psline{-}(4.7,0.3)(1.5,1.6) \psline{-}(4.7,0.3)(0.7,2.4)
\psline{-}(4.7,0.3)(1.1,3.3) \psline{-}(4.7,0.3)(0.8,1.2)
\psline{-}(4.8,3.6)(1.5,1.6) \psline{-}(4.8,3.6)(0.7,2.4)
\psline{-}(4.8,3.6)(1.1,3.3) \psline{-}(4.8,3.6)(0.8,1.2)
\psline{-}(5,1.5)(1.5,1.6) \psline{-}(5,1.5)(0.7,2.4)
\psline{-}(5,1.5)(1.1,3.3) \psline{-}(5,1.5)(0.8,1.2)
\psellipse[linestyle=dashed, fillstyle=none](1,2)(1,2)
\uput[u](8.8,8){\large{\textbf{t}}}
\uput[u](8.8,2){\large{\textbf{b}}}
\uput[u](5.5,9.7){\large{\textbf{$I_k$}}}
\uput[u](5.5,3.7){\large{\textbf{$I_{k+1}$}}}
\uput[u](1.5,9.7){\large{\textbf{$G_1$}}}
\uput[u](1.5,3.7){\large{\textbf{$G_2$}}}
\end{pspicture}
}

\begin{figure}[h]
  \centering
\includegraphics[width=5cm]{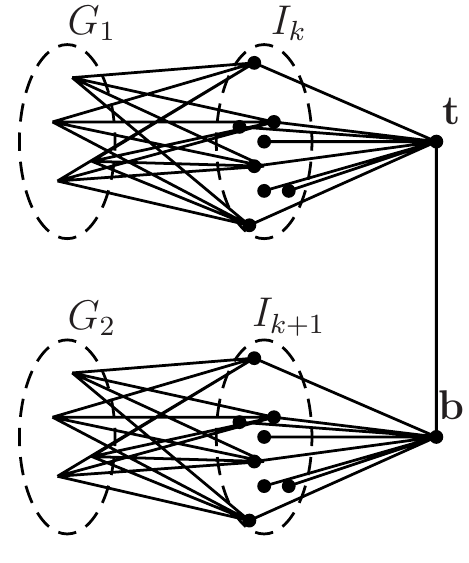}
  \caption{ \ The block
$B_k(G,t)$}\label{fig:fig1}\vspace{-3mm}
\end{figure}

 In order to obtain hardness for CQA under
the C-repair semantics, we need to construct the block graph
$B_k(G,{\bf t})$ (c.f. Figure 1), consisting of two copies
$G_1,G_2$ of $G$, and two internally disconnected subgraphs $I_k,
I_{k+1}$, with $k$ and $k+1$ vertices, resp. Every vertex in $G$
($G'$) is connected to every vertex in $I_k$ (resp. $I_{k+1}$).

\begin{lemma} \label{lem:block} \em ~ Given a  graph
$G$ and a number $k$,  a graph $B_k(G,{\bf t})$ can be computed in
polynomial time in the size of $G$, where ${\bf t}$ is a
distinguished vertex in it that belongs to all its maximum
independent sets iff the cardinality of a maximum independent set
of $G$ is equal to $k$. 
\end{lemma}

\vspace{-1mm}


\dproof{Lemma \ref{lem:block}}{
The new graph $G'$ consists of two copies of $G$, say $G_1, G_2$,
two additional graphs, $I_k, I_{k+1}$, and two extra vertices
$t,b$. Subgraph $I_k$ consists of $k$ mutually disconnected
vertices; subgraph $I_{k+1}$ consists of $k+1$ mutually
disconnected connected vertexes. Each vertex of $G_1$ is adjacent
to each vertex of $I_k$, and each vertex of $G_2$ is adjacent to
each vertex of $I_{k+1}$. Each vertex of $I_k$ is adjacent to $t$,
and each vertex of $I_{k+1}$ is adjacent to $b$.  Finally, $t, b$
are connected by an edge (c.f. Figure 1).

 We claim that vertex $t$ belongs to all maximum independent sets
of $G'$ iff the cardinality of maximum independent set of $G$ is
equal to $k$. To prove this claim, we consider a few, but
representative possible cases. With $I(G)$ we denote an arbitrary
maximum independent set of $G$.
\begin{itemize}
 \item [1.] $|I(G)| < k - 1$: The maximum independent set of $G'$ is
 $I_k \cup I_{k+1}$; with cardinality
 $2k + 1$.
 \item [2.] $|I(G)| = k - 1$: The maximum independent sets of $G'$
 are (a) $I(G_1) \cup I_{k+1} \cup
 \{t\}$, and
 (b) $I_k \cup I_{k+1}$, with cardinality $2k+1$.
 \item [3.] $|I(G)| = k$: The maximum independent set of $G'$ is
 $I_{k+1} \cup I(G_1) \cup \{t\}$, with cardinality $2k+2$.
 \item [4.] $|I(G)| = k + 1$: The maximum independent sets of $G'$ are
 (a) $G_1 \cup G_2 \cup \{t\}$,  (b) $G_1 \cup G_2
 \cup \{b\}$, (c) $G_1 \cup I_{k+1} \cup \{t\}$; with cardinality $2k+3$.
 \item [5.] $|I(G)| > k + 1$: The maximum independent sets of $G'$ are
 (a) $G_1 \cup G_2 \cup \{t\}$, (b) $G_1 \cup G_2 \cup
\{ b\}$;
 with cardinality $2|I| +1$.
\end{itemize}
Only in case $|I(G)| = k$, $t$ belongs to all maximum independent
sets.}

\begin{lemma} \label{prop:allsets} \em Deciding if a
vertex belongs to all maximum independent sets of a graph is
$P^{\nit{NP(log}( n))}$-hard. \boxtheorem
\end{lemma}
This result can be proved by reduction from the following
$P^{\nit{NP}(\nit{log}(n))}$-complete decision problem
\cite{krentel}: Given a graph $G$ and an integer $k$, is the size
of a maximum clique in $G$ equivalent to $0~ \nit{mod}~ k$? $G$ is
reduced to a graph $G'$ that is built by combining a number of
versions of the block construction in Figure 1.

\vspace{3mm}
\dproof{Lemma \ref{prop:allsets}}{
By reduction from the following
$P^{\nit{NP}(\nit{log}(n))}$-complete decision problem
\cite[theorem 3.5]{krentel}: Given a graph $G$ and an integer $k$,
is the size of a maximum clique in $G$ equivalent to $0~
\nit{mod}~ k$?

Assume graph $G$ has $n$ vertices. We can also assume that $k$ is
not bigger than $n$. Now, we pass to the graph $G'$ that is the
complement of $G$: It has the same vertices as $G$, with every two
distinct vertices being adjacent in $G'$ iff they are not adjacent
in $G$. A maximum independent set of $G'$ is a maximum clique of
$G$ and vice-versa. So, the cardinality of a maximum independent
set of $G'$ is the size of a maximum clique of $G$.

Next, we take advantage of the construction in Lemma
\ref{lem:block} (c.f. Figure 1): For each $m \in \{k, 2k, \cdots,
\lfloor \frac{n}{k} \times k\rfloor \}$, construct the block graph
$B_m(G', t_m)$. (There are $[n/k]$ possible solutions to the
equation $x \equiv 0 ~\nit{mod}~k$.) All these graphs are
disconnected from each other. Next, create a new vertex $t_g$ and
connect it to the vertices $t_m$  of the blocks $B_m(G',t_m)$. It
is easy to check that the resulting graph, say $\overline{G}$, has
its size bounded above by $O(n^4)$.

It holds that vertex $t_g$ does not belong to every maximum
independent set of $\overline{G}$ iff the size of maximum
independent set of $G$ is equivalent to $0\ \nit{mod}\ k$. So, we
have a reduction to the complement of our problem, but the class
$P^{\nit{NP(log}(n))}$ is closed under complement.

In fact, if the size of maximum independent set of $G$ is not
equivalent to $0~ \nit{mod}~k$, then for every block $B$ in
$\overline{G}$, there exists a maximum independent set $I_B$ of
the block $B$ such that $t_B \notin I_B$ ($t_B$ is the top node of
block $B$). The maximum independent set of $\overline{G}$ is
$\{t_g\} \cup \bigcup_B I_B$ (because there are no edges between
blocks and between $t_g$ and other vertices besides $t_B$).
Consider any independent set $I$ of $\overline{G}$ that does not
contain $t_g$. The size of the projection of $I$ on any block is
not greater than the size of the maximum independent set of the
block; so $|I| \leq |\bigcup_B I_B|$. So, $t_g$ belongs to every
maximum independent set of $\overline{G}$.

Now, if the size of a maximum independent  set  of $G$ is
equivalent to $0\ \nit{mod}\ k$, then there exists one block
$B_{\!o}$ such that $t_{B_o}$ belongs to every maximum independent
set $I_{B_o}$ of $B_{\!o}$, while for all other blocks $B$ there
exists $I_B$ such that $t_B \notin I_B$. Consider a maximum
independent set $I_t$ of $\overline{G}$ that contains $t_G$.

Every maximum independent set of $\overline{G}$ that contains
$t_g$ is of the form $\{t_g\}$ union of maximum independent sets
from the blocks $B$ other than $B_o$ that do not contain their
corresponding $t_B$ union any maximum independent set of $B_{\!o}
\smallsetminus \{t_{B_o}\}$. The size of such a set is $s = 1 +
\sum_{B \neq B_o} |I(B)| + (|I_{B_o}| -1)$. A maximum independent
set $I$ that does not contain $t_g$, is the union of maximum
independent sets $I_B$ of all the blocks $B$ of $\overline{G}$,
and its size is equal to $\sum_B |I_B|$, i.e. $s$. Then, there
exists a maximum independent set that does not contain $t_g$. }\\

Now, the graph
$G'$ used in Lemma \ref{prop:allsets} can be represented according
to Lemma \ref{lem:everyGraph} as a database consistency problem,
and in this way we obtain

\begin{theorem} \label{teo:hardness} \em
For denial constraints, CQA under the C-repair semantics for
queries that are conjunctions of ground literals  is
$P^{\nit{NP}(\nit{log}( n))}$-complete. 
\end{theorem}

\dproof{Theorem \ref{teo:hardness}}{Membership follows from
Theorem \ref{prop:membDCs}. Now we prove hardness.
 For a graph $G$ and integer $k$, we
construct a database $D$, such that the consistent answer to a
ground atomic query $Q$ can be used to decide if the size of a
maximum clique of $G$ is equivalent to $0~ \nit{mod}~ k$ (c.f.
proof of Lemma \ref{prop:allsets}). Construct the graph
$\overline{G}$ as in Lemma \ref{prop:allsets}. Encode graph
$\overline{G}$ as a database inconsistency problem, introducing a
unary relation $V$ (for vertices) and $E$ (3-ary), where $E$
corresponds to the edge relation in $\overline{G}$ plus a third
padding attribute to make changing it more costly. For each vertex
$v \in \overline{G}$, there is a tuple $(v)$ in $V$.

We also introduce the denial constraint: $\forall v_1 \forall v_2
\neg(V(v_1) \wedge V(v_2) \wedge E(v_1, v_2,\!\_))$ (an underscore
means any variable implicitly universally quantified).
 For each edge
$\{v_1, v_2\} \in \overline{G}$, create $n$ different versions
$(v_1, v_2,p)$ in $E$, as in the proof of Lemma
\ref{lem:everyGraph}. The effect of fixing the database wrt the
given denial constraint may be the removal of tuples representing
vertices or/and the removal of  tuples representing edges. We want
to forbid the latter alternative because those repairs do not
represent maximum independent set; and this is achieved by making
them more expensive than vertex removal through the padding
process.

The consistent answer to the query $V(t_g)$ is {\em no}, i.e. not
true in all repairs, iff $t_g$ does not belong to all maximum
independent sets of $\overline{G}$ iff the size of a maximum
independent set of $G'$ is equivalent to $0~ \nit{mod}~ k$ iff the
size of a maximum clique of $G$ is equivalent to $0~ \nit{mod}~
k$.}\\

This theorem still holds for ground atomic queries, which is
interesting, because for this kind of queries and denial
constraints CQA under the S-repair semantics is in \textit{PTIME}
\cite{cm-2005}. \ignore{It is also worth observing that query
answering under the S-repair semantics in the context of belief
revision/update is more complex than the same problem for the
C-repair semantics (assuming the polynomial hierarchy does not
collapse); more precisely Winslett's framework \cite{winslett}
(based on set inclusion) is $\Pi^P_2$-complete, while Dalal's
\cite{dalal} (based on set cardinality) is
$P^\nit{NP(log(n))}$-complete \cite{eg-92}. However, these results
for belief revision/update are not about data complexity. C.f.
\cite{abc-1999} for a discussion around CQA vs. belief
revision/update.}

\section{Incremental Complexity of CQA}\label{sec:incr}

Assume that we have a consistent database instance $D$ wrt to
$\IC$. $D$ may become inconsistent after the execution of an
update sequence $U$ composed of operations of the forms
$\nit{insert}(R(\bar{t})), ~\nit{delete}(R(\bar{t}))$, ~meaning
insert/delete tuple $R(\bar{t})$ into/from $D$, ~or
$\nit{change}(R(\bar{t}),A,a)$, for changing value of attribute
$A$ in $R(\bar{t})$ to $a$, with $a \in {\cal U}$.
 We are interested in  whether we can
find consistent query answers from the possibly inconsistently
updated database $U(D)$ more efficiently by taking into account
the previous consistent database state.
\begin{definition} \em For  a consistent
  database $D$ wrt $\IC$,
  and a sequence $U$ of update operations $U_1,\ldots,U_m$,
\emph{incremental consistent query answering} for query $Q$ is CQA
for $Q$ wrt $\IC$  from  instance
  $U(D)$, that results
from applying $U$ to $D$. \boxtheorem
\end{definition}
Update sequences $U$ will be atomic, in the sense that they are
completely executed or not. This allows us to concentrate on
``minimized" versions of update sequences, e.g. containing only
insertions and/or attribute changes when dealing with denial
constraints, because deletions do not cause any violations. We are
still interested in data complexity, i.e. wrt the size $|D|$ of
the original database. In particular, $m$ is fixed, and usually
small wrt $|D|$.

A notion of incremental complexity has been introduced in
\cite{msvt-1994}, and also in \cite{imm-99} under the name of {\em
dynamic complexity}. There, the instance that is updated can be
arbitrary, and the question is about the complexity for the
updated version when information about the previous instance can
be used. In our case, we are assuming that the initial database is
consistent. As opposed to \cite{msvt-1994,imm-99}, where new
incremental or dynamic complexity classes are introduced, we
appeal to those classic complexity classes found at a low level in
the polynomial hierarchy.

\subsection{Incremental complexity: C-repair semantics}
\label{sec:incmcss}

In contrast to static CQA for the C-repair semantics, it holds

\begin{theorem}\label{prop:incr-dens}
\em For the C-repair semantics, first-order boolean queries,
denial constraints, and update
  sequences $U$ of fixed length $m$ applied to $D$, incremental
  CQA
   is in \textit{PTIME} in $|D|$. 
  \end{theorem}

  \dproof{Theorem \ref{prop:incr-dens}}{
For denial constraints tuple deletions do not introduce any
violations, so we consider a sequence $U$ consisting of tuple
insertion and updates.

 Assume that $k$ of the $m$ inserted
tuples violate ICs, perhaps together with some tuples already in
$D$. If we delete $k$ violating tuples, then we get a consistent
database $D'$;  so a minimal repair is at a distance less than or
equal to $k$ from $D$. To find all minimal repairs it is good
enough to check no more than $N = {\small \left(
\begin{array}{c} n + m \\ 1 \end{array} \right) + \left(
\begin{array}{c} n + m\\ 2
\end{array} \right) + \cdots\ + \left( \begin{array}{c} n + m\\ k
\end{array} \right)}$
 repairs, where $|D| = n$. If $m$ is small, say less than $c
\cdot n$, then $N < k {\small \left( \begin{array}{c} n + m\\
k\end{array}\right)}  \leq m {\small \left( \begin{array}{c} n\\
m\end{array}\right)}^{\!\!m} <  mn^m$. Thus, the incremental
complexity of the CQA is polynomial wrt $n$.

In case $U$ contains change updates, the proof is essentially the
same, but the role of $m$ is taken by $m \cdot a$, where $a$ is
the maximum arity of the relations involved. This is because we
have to consider possible changes in different attributes.}\\

The proof of this theorem provides  an upper bound of of $O(m
\cdot n^m)$, that is polynomial in the size $n$ of the initial
database, but exponential in $m$, which makes the problem
tractable in data, but with the size of the update sequence in the
exponent. We are interested in determining if queries can be
consistently answered in time $O(f(m) \times n^c)$, for a constant
$c$ and a function $f(m)$ depending only on $m$. In this way we
isolate the complexity introduced by $U$.

The area of parameterized complexity studies this kind of problems
\cite{grohe,papyan-1999}. A decision problem with inputs of the
form $(I,p)$, where $p$ ~is a distinguished parameter of the
input, is {\em fixed parameter tractable}, and by definition
belongs to the class $\nit{FPT}$ \cite{downfel}, if it can be
solved in time $O(f(|p|) \cdot |I|^c)$, where $c$ and the hidden
constant do not depend on $|p|$ or $|I|$ and $f$ does not depend
on $|I|$.

\begin{definition}  ~\em Given a query $Q$,
ICs $\IC$, and a ground tuple $\bar{t}$, \emph{parameterized
incremental CQA} is the decision problem $\nit{CQA}^p( Q,$ $\IC)
:= \{(D,U,\bar{t}) ~|~ D$ $\mbox{is an}$ $\mbox{instance, } U
\mbox{ an}$ $\mbox{update sequence },$ $\bar{t}~ \mbox{ is
consistent answer to } Q \mbox{ in } U(D)\}$, whose parameter is
$U$, and  consistency  of answers refers to C-repairs of $U(D)$.
\boxtheorem
\end{definition}
We keep $Q$ and $\IC$ fixed in the problem definition because,
except for the parameter $U$, we are interested in data
complexity.

\begin{theorem}\label{prop:fpt} \em
For functional dependencies and queries that are conjunctions of
literals,  parameterized incremental CQA is in $\nit{FPT}$.
\end{theorem}

\dproof{Theorem \ref{prop:fpt}}{
First, it is known that the problem of, given a graph $G$ and a
number $k$, determining if there exists a vertex cover of size
less than or equal to $k$ is in FPT \cite{downfel}. We will use
this problem to solve ours.

Now, let us assume that we have a consistent database $D$ of size
$n$, and we update it inserting $k$ new tuples, obtaining an
inconsistent database $D'$ with conflict graph $G$. The size of
$G$ is $O(n)$ by our assumption on the size of $m$ in comparison
with $n$. Every C-repair of $D'$ is a maximum independent set of
$G$, and can be obtained by deleting from $G$ a minimum vertex
cover, because the problems are complementary. So, a minimum
vertex cover corresponds to the vertices that are to be deleted to
obtain a repair.

Since the original database $D$ is consistent, the vertices of $G$
corresponding to database tuples in $D$ are all disconnected from
each other. In consequence, edges may appear only by the update
sequence, namely between the $m$ new tuples or between them and
the elements of $D$. Then, we know that there is a vertex cover
for $G$ of size $m$. However, we  do not know if it is minimum.

In order to find the size of a minimum vertex cover of $G$, we may
start doing binary search from $m$, applying an FPT algorithm for
vertex cover.  Each check for vertex cover, say for value $m_i$,
can be done in $O(1.2852^m_i + m_i \cdot n)$ \cite{chen}. Then
$\nit{log}(m)$ checks take time $O(\nit{log}(m) \cdot (1.2852^m +
m \cdot n)) \leq O(f(m) \cdot n)$, with $f$ an exponential
function in $m$. So, it is in $\nit{FPT}$ obtaining the size of a
minimum vertex cover for $G$, which gives us the minimum number of
tuples to remove to restore consistency.

Now, for CQA we want to check if a vertex $R(\bar{t})$ belongs to
all maximum independent sets of $G$, which happens if it does not
belong to any minimum vertex covers. This can be determined by
checking the size of minimum vertex cover for $G'$ and $G'
\smallsetminus \{R(\bar{t})\}$. If they are the same, then
$R(\bar{t})$ belongs to all maximum independent sets and the
consistent answer to the query $R(\bar{t})$ is $\nit{yes}$.}\\

The \emph{vertex cover problem}, of  deciding if graph $G$ has a
vertex cover (VC) of size no bigger than $k$, belongs to the class
$\nit{FPT}$, i.e. there is a polynomial time parameterized
algorithm $\nit{VC(G,k)}$ for it \cite{downfel}; actually one that
runs in time $O(1.2852^k + k \cdot n)$, being $n$ the size of $G$
\cite{chen}.

The algorithm whose existence is claimed in Theorem \ref{prop:fpt}
is  as follows: Let $G$ be the conflict graph associated to the
database obtained after the insertion of $m$ tuples. By binary
search, calling each time $\nit{VC}(G,\_)$, it is possible to
determine the size of a minimum VC for $G$. This gives us the
minimum number of tuples that have to be removed in order to
restore consistency; and can be done in time $O(\nit{log}(m) \cdot
(1.2852^m + m \cdot n))$, where $n$ is the size of the original
database. In order to determine if a tuple $R(\bar{t})$ belongs to
every maximum independent set, i.e. if it is consistently true,
compute the size of a minimum VC for $G \smallsetminus
\{R(\bar{t})\}$. The two numbers are the same iff the answer is
$\nit{yes}$. The total time is still $O(\nit{log}(m) \cdot
(1.2852^m + m \cdot n)))$, which is linear in the size of the
original database. The same algorithm applies if, in addition to
tuple insertions, we also have changes of attribute values in the
update part; of course, still under the C-repair semantics.

Theorem \ref{prop:fpt} uses the membership to $\nit{FPT}$ of the
VC problem, which we apply to conflict graphs for functional
dependencies. However, the result can be extended to denials
constraints and their conflict hypergraphs. In our case, the
maximum size of an hyperedge is the maximum number of database
atoms in a denial constraint, which is determined by the fixed
database schema. If this number is $d$, then we are in the
presence of the so-called \emph{d-hitting set problem}, consisting
in finding the size of a minimum hitting set for an hypergraph
with hyperedges bounded in size by $d$. This problem is in
$\nit{FPT}$ \cite{niedermeier}.

\begin{theorem} \label{teo:fpt} \em
For denial constrains and queries that are conjunctions of
literals,  parameterized incremental CQA is in $\nit{FPT}$.
\boxtheorem
\end{theorem}
Using the reductions in Section \ref{sec:mcss}, this result can be
extended to incremental CQA under the {\em possible} C-repair
semantics.

\subsection{Incremental complexity: S-repair semantics}\label{sec:s}
Incremental CQA for non-quantified conjunctive queries under
denial constraints belongs to $\nit{PTIME}$, which can be
established by applying the algorithm in \cite{cm-2005} for the
static case to $U(D)$.

However,  for quantified conjunctive queries the situation may
change. Actually, by reduction from static
  CQA for conjunctive queries and denial ICs under the S-repair semantics,
  which is \textit{coNP}-hard
  \cite{cm-2005}, we obtain
\begin{theorem}\label{lem:incr-icd-tds-msd} \em
Under the S-repair semantics,
  incremental CQA for conjunctive queries and denial constraints is
  \textit{coNP}-hard. 
\end{theorem}

\dproof{Theorem \ref{lem:incr-icd-tds-msd}}{By reduction from static
  CQA for (existentially quantified) conjunctive queries and denial ICs under
  minimal set semantics, which is \textit{coNP}-hard
  \cite{cm-2005}. Consider an instance for this problem consisting of a database
  $D$, a set of denial ICs $\IC$, and a query $Q$.

  For every denial $\nit{ic} \in \IC$, pick up a relation $R^\nit{ic}$ in it and
  expand it to a relation
  $\overline{R^\nit{ic}}$ with an extra attribute $\nit{Control}$.
  Also add a new, one attribute relation
  $\nit{Controler}(A)$. Next, transform each integrity constraint
$\nit{ic}\!: ~
  \forall \bar{x} \neg (P(\bar{x}) \land \cdots \wedge R^\nit{ic}(\bar{x}) \land
  \cdots \wedge \gamma)$ into
  $\nit{ic}^{\prime}\!: ~\forall \bar{x}\forall \nit{contr} \neg (P(\bar{x}) \land
  \cdots \wedge
  \overline{R^\nit{ic}}(\bar{x},
  \nit{contr}) \land \nit{Controler(contr)} \land \gamma)$. We
  obtain a set $\IC'$ of denial constraints. The original database
  $D$ is extended to a database $\overline{D}$ with the new relation
  $\nit{Controler}$, which is initially empty, and the relations $\overline{R^\nit{ic}}$,
  whose extra attributes $\nit{Contr}$ initially take all
the value $1$. Due to the extension of $\nit{Controler}$, $\IC'$
is satisfied.

  Now in the incremental context, we consider the inconsistent instance $\overline{D}'$
  obtained via the update $\nit{insert}(\nit{Controler}(1))$ on $\overline{D}$.
The S-repairs  of $\overline{D}'$ wrt $\IC'$ are: (a)
$\overline{D}$ and (b) all the S-repairs
  of $\overline{D}$ (plus the tuple $\nit{Controler}(1)$ in each of them), which are in one-to-one
  correspondence with the S-repairs of $D$ wrt $\IC$.
Now, for a conjunctive query $Q$ in the language of $D$, produce
the conjunctive query
  $Q'\!: \exists \cdots y_\nit{ic} \cdots Q\frac{\cdots R^\nit{ic}(\bar{x}) \cdots}{\vspace*{1mm}\cdots
  \overline{R^\nit{ic}}(\bar{x},y_\nit{ic}) \cdots}$ in the
  language of $\overline{D}$,\footnote{$E\frac{E_1}{E_2}$ means the expression obtained by replacing
  in expression $E$ the subexpression $E_1$ by expression $E_2$.} where each atom
  $R^\nit{ic}(\bar{x}))$ in $Q$ is replaced by $\exists y_\nit{ic}
  \overline{R^\nit{ic}}(\bar{x},y_\nit{ic})$.

  Notice that all the repairs in (b) are essentially contained
 in $\overline{D}$, except for the tuple $\nit{Controler}(1)$, whose
  predicate does not appear in the queries. This is because denial
  constraints are obtained by tuple deletions. In consequence, any
  answer to the conjunctive (and then monotone) query in a repair
  in (b) is also an answer in the repair in (a). In consequence,
  the repair $\overline{D}$ does not contribute with any new
  consistent answers, neither invalidates any answers obtained by
  the repairs in (b). So, it holds $\nit{Cqa}(Q,D,\IC) =
  \nit{Cqa}(Q',\overline{D}',\IC')$.}\\

We can see that, for denial constraints, static CQA under the
C-repair semantics seems to be harder than under the S-repair
semantics ($P^{\nit{NP(log}(n))}$- vs. $\nit{coNP}$-hard). On the
other side, incremental CQA under the S-repair semantics seems to
harder than under the C-repair semantics ($\nit{coNP}$-hard vs.
\nit{PTIME}). The reason is that for the C-repair semantics the
cost of a repair cannot exceed the size of the update, whereas for
the S-repair semantics the cost of a repair may be unbounded wrt
the size of an update.

\begin{example}\label{ex:comp}
Consider a schema $R(\cdot), S(\cdot)$ with the denial constraint
$\forall x \forall y\neg (R(x) \wedge S(y))$; and the consistent
database $D = \{R(1), \ldots, R(n)\}$, with an empty table for
$S$. After the update $U= insert(S(0))$, the database becomes
inconsistent, and the S-repairs are $\{R(1), \ldots, R(n)\}$ and
$\{S(0)\}$. However, only the former is a C-repair, and is at a
distance $1$ from the original instance, i.e. as the size of the
update. However, the second S-repair is at a distance $n$.
\boxtheorem
\end{example}

\subsection{Incremental complexity: A-repair semantics}
\label{sec:attrib}

Before addressing the problem of incremental complexity, we give a
complexity lower bound for the {\em weighted} version of static
CQA for the A-repair semantics.  In this case, we have a numerical
weight function $w$ defined on triples of the form
$(R(\bar{t}),A,\nit{newValue})$, where $R(\bar{t})$ is a database
tuple stored in the database, $A$ is an attribute of $R$, and
$\nit{newValue}$ is a new value for $A$ in $R(\bar{t})$. The
\emph{weighted A-repair semantics} (wA-repair semantics) is just a
particular case of Definition \ref{def:dist}(c), where the
distance is given by an aggregation function $g$ applied to the
set of numbers $\{w(R(\bar{t}),A,\nit{newValue})~|~ R(\bar{t}) \in
D\}$.

Typically, $g$~ is the sum, and the weights are
$w(R(\bar{t}),A,\nit{newValue}) = 1$ if $R(\bar{t})[A]$ is
different from $\nit{newValue}$, and $0$ otherwise, where
$R(\bar{t})[A]$ is the projection of database tuple $R(\bar{t})$
on attribute $A$, i.e. just the number of changes is counted
\cite{franconi}. In \cite{bbfl-2005}, $g$ is still the sum, but
$w$ is given by $w(R(\bar{t}),A,\nit{newValue}) = \alpha_{\!A}
\!\cdot \!(R(\bar{t})[A] - \nit{newValue})^2$, where $\alpha_A$ is
a coefficient introduced to capture the  relative importance of
attribute $A$ or scale factors. In these cases, $w$ does not
depend on $D$. However, if the weight function $w$ depended on the
size of $D$, $w$ should become part of the input for the decision
problem of CQA.

\begin{theorem}\label{theo:wmccs} \em Static CQA for ground
atomic queries and denial constraints under the wA-repair
semantics is $P^\nit{NP}$-hard. 
\end{theorem}

  \dproof{Theorem \ref{theo:wmccs}}{ We provide a $\nit{LOGSPACE}$-reduction
from the following problem \cite[theorem 3.4]{krentel}: Given a
Boolean formula $\psi(X_1, \cdots, X_n)$ in 3CNF, decide if the
last variable $X_n$ is equal to $1$ in the lexicographically
maximum satisfying assignment (the answer is ${\it No}$ if $\psi$
is not satisfiable).

Create a database schema with relations:~ $\nit{Clause(id,
Var}_1,$ $ \nit{Val}_1, \nit{Var}_2, \nit{Val}_2,$ $\nit{Var}_3,
\nit{Val}_3)$, $\nit{Var(var, val)}$, $\nit{Dummy}(x)$, with
denial constraints:\\
$\forall var, val  \neg (Var(var, val) \land val \not = 0  \land
val \not = 1)$,\\
$\forall id, v_1, x_1, v_2, x_2, v_3, x_3 \neg (Cl(id, v_1, x_1,
v_2, x_2, v_3, x_3)
 \land Var(\_, v_1, x'_1)  \land Var(\_, v_2, x'_2)$
 $\land~ Var(\_, v_3, x'_3) ~\land~
 x_1 \not = x'_1  \land x_2 \not = x'_2 \land x_3 \not =  x'_3
\land
 \nit{Dummy}(1))$.

\noindent The last denial can be replaced by 8 denial constraints
without inequalities considering all the combination of values for
$x_1, x_2, x_3$ in $\{0,1\}$.

Assume now that $C_1, \ldots, C_m$ are the clauses in $\psi$. For
each propositional variable $X_i$ store in table $\nit{Var}$ the
tuple $(X_i, 0)$, with weight $1$, and $(X_i,1)$ with weight
$2^{n-i}$. Store tuple $1$ in $Dummy$ with weight $2^n\times 2$.
For each clause $C_i = l_{i_1} \lor l_{i_2} \lor l_{i_3}$, store
in $\nit{Clause}$  the tuple $(C_i, X_{i_1}, \tilde{l}_{i_1},
X_{i_2}, \tilde{l}_{i_2}, X_{i_3}, \tilde{l}_{i_3})$, where
$\tilde{l}_{i_j}$ is equal to $1$ in case of positive occurrence
of variable $X_{i_j}$ in $C_i$; and  to $0$, otherwise. For
example, for $C_6 = X_6 \lor \neg X_9 \lor X_{12}$, we store
$(C_6, X_6,1, X_9, 0,  X_{12}, 1)$. The weight of this tuple is
$2^n$.

Then the answer to the ground atomic query $\nit{Var}(X_i,1)$ is
$\nit{yes}$ iff the variable $X_i$ is assigned value $1$ in the
lexicographically maximum assignment (in case  such a satisfying
assignment exists). In case a satisfying assignment does not
exist, then the tuple in $\nit{Dummy}$ has to be changed in order
to satisfy the constraints. No attribute value in a tuple in
$\nit{Clause}$ is changed, because the cost of such a change is
higher than a change in the $\nit{Dummy}$ relation.}\\

In order to obtain a hardness result in the incremental case and
for denial constraints (for which we are assuming update sequences
do not contain tuple deletions), we can use the kind of A-repairs
introduced in \cite{bbfl-2005}.

\begin{theorem}\label{prop:incr-attr-denial} \em Incremental
CQA for atomic queries and denial constraints under the wA-repair
semantics is $P^\nit{NP}$-hard. 
\end{theorem}

\dproof{Theorem \ref{prop:incr-attr-denial}}{
By reduction from the problem $P$ of CQA in \cite[theorem
4(b)]{bbfl-2005}. We introduce a new relation $\nit{Dummy}$, and
transform every denial $\forall \bar{y} \neg (A_1 \wedge \cdots
\wedge A_s)$ for  problem $P$ into $\forall \bar{y} \forall x \neg
(A_1 \wedge \cdots \wedge A_s \wedge \nit{Dummy}(x))$.
 If we start with the empty extension for $\nit{Dummy}$, the
database is consistent. On the update part, if we insert the tuple
$\nit{Dummy}(c)$ into the database, and the original denials were
inconsistent in the given instance, then we cannot delete that
tuple and no change in it can repair any violations. Thus, the
only way to repair database is  as in \cite{bbfl-2005}, which
makes CQA $P^\nit{NP}$-hard.}\\

These results still hold  for tuple insertions as update
actions, the fixed weight function that assigns value $1$ to every
change, and the sum as aggregation function. In case we have
numerical values as in \cite{bbfl-2005} or a bounded domain, we
can obtain as in \cite[theorem 4(b)]{bbfl-2005} that the problems
in Theorems \ref{theo:wmccs} and \ref{prop:incr-attr-denial}
belong both to $\Pi^P_2$.

 Under the A-repair semantics,
if the update sequence consist of $\nit{change}$ actions, then we
can obtain polynomial time incremental CQA under the additional
condition that the set of attribute values than can be used to
restore consistency is bounded in size, independently from the
database (or its active domain). Such an assumption can be
justified in several applications, like in census-like databases
that are corrected according to inequality-free denial constraints
that force the new values to be taken at the border of a database
independent region
 \cite{bbfl-2005}; and also in applications where denial constraints, this time
containing inequalities, force the attribute values to be taken in
a finite, pre-specified set. The proof is similar to that of
Theorem \ref{prop:incr-dens}, and the polynomial bound now also
depends on the size of the set of candidate values.

\begin{theorem}\label{prop:bounded} \em
For a database independent and bounded domain of attribute values,
 incremental
  CQA  under the A-repair semantics, for first-order boolean queries, denial constraints, and
  update sequences containing
  only $\nit{change}$ actions is in \textit{PTIME} in the size of the original database.
\boxtheorem
\end{theorem}
Now, we present a lower bound for CQA under the A-repair semantics
for {\em first-order ICs} and tuple deletions, which now may
affect their satisfaction.
\begin{lemma}\label{lem:3col4reg} \em For any planar graph $G$
with vertices of degree at most 4, there exists a regular graph
$G'$ of degree 4 that is 4-colorable, such that $G'$ is
3-colorable iff $G$ is 3-colorable. $G'$ can be built in
polynomial time in $|G|$. 
\end{lemma}

\dproof{Lemma \ref{lem:3col4reg}}{If a vertex $v$ in $G$ has degree 2, then
we transform it into a vertex of degree 4 by hanging from it an
``ear" as shown in the figure, which is composed of three
connected versions of the graph $H_3$ \cite[Theorem 2.3]{gjs-1976}
plus two interconnected versions of a box graph (c.f. Figure \ref{fig:fig2}).

\vspace{-2mm}
\ignore{
\begin{multicols}{2}
\vspace*{-6mm}
\psset{xunit=1mm,yunit=1mm}
\begin{pspicture}(-5,-8)(55,60) \psline(25,50)(28,58)
\psline(25,50)(22,58) 
\uput[r](25,50){$v$~~r} \psdot(20,40) \uput[r](20,40){g}
\psdot(30,40) \uput[r](30,40){b} \psdot(25,35) \uput[r](25,35){r}
\psdot(15,30) \uput[d](15,30){r} \psdot(21,30) \uput[d](21,30){b}
\psdot(29,30) \uput[d](29,30){g} \psdot(35,30) \uput[r](35,30){r}
\pscircle(15,30){5pt} \pscircle(35,30){5pt} \psline(25,50)(20,40)
\psline(25,50)(30,40) \psline(20,40)(25,35) \psline(25,35)(30,40)
\psline(20,40)(15,30) \psline(20,40)(21,30) \psline(25,35)(21,30)
\psline(25,35)(29,30) \psline(30,40)(29,30) \psline(30,40)(35,30)
\psline(15,30)(21,30) \psline(21,30)(29,30) \psline(29,30)(35,30)
\psdot(15,30) \psdot(10,20) \uput[r](10,20){g} \psdot(20,20)
\uput[r](20,20){b} \psdot(15,15) \uput[r](15,15){r} \psdot(5,10)
\uput[u](5,10){r} \psdot(11,10) \uput[u](11,10){b} \psdot(19,10)
\uput[u](19,10){g} \psdot(23,10) \uput[u](23,10){r}
\psline(15,30)(10,20) \psline(15,30)(20,20) \psline(10,20)(15,15)
\psline(15,15)(20,20) \psline(10,20)(5,10) \psline(10,20)(11,10)
\psline(15,15)(11,10) \psline(15,15)(19,10) \psline(20,20)(19,10)
\psline(20,20)(23,10) \psline(5,10)(11,10) \psline(11,10)(19,10)
\psline(19,10)(23,10) \psdot(35,30) \psdot(30,20)
\uput[r](30,20){g} \psdot(40,20) \uput[r](40,20){b} \psdot(35,15)
\uput[r](35,15){r} \psdot(27,10) \uput[u](27,10){r} \psdot(31,10)
\uput[u](31,20){b} \psdot(39,10) \uput[u](39,10){g} \psdot(45,10)
\uput[u](45,10){r} \psline(35,30)(30,20) \psline(35,30)(40,20)
\psline(30,20)(35,15) \psline(35,15)(40,20) \psline(30,20)(27,10)
\psline(30,20)(31,10) \psline(35,15)(31,10) \psline(35,15)(39,10)
\psline(40,20)(39,10) \psline(40,20)(45,10) \psline(27,10)(31,10)
\psline(31,10)(39,10) \psline(39,10)(45,10) \psdot(5,5)
\uput[d](5,5){b} \psdot(15,5) \uput[d](15,5){g} \psdot(23,5)
\uput[d](23,5){b} \psline(5,5)(15,5) \psline(15,5)(23,5)
\psline(15,5)(5,10) \psline(15,5)(23,10) \psline(5,5)(5,10)
\psline(23,5)(23,10) \psdot(27,5) \uput[d](27,5){g} \psdot(35,5)
\uput[d](35,5){b} \psdot(45,5) \uput[d](45,5){g}
\psline(27,5)(35,5) \psline(35,5)(45,5) \psline(35,5)(27,10)
\psline(35,5)(45,10) \psline(27,5)(27,10) \psline(45,5)(45,10)
\pscurve(5,5)(15,1)(27,5) \pscurve(5,5)(25,1)(45,5)
\psline(23,5)(27,5) \pscurve(23,5)(35,1)(45,5)
\end{pspicture}
}

\begin{figure}
  \centering
  \includegraphics[width=5cm]{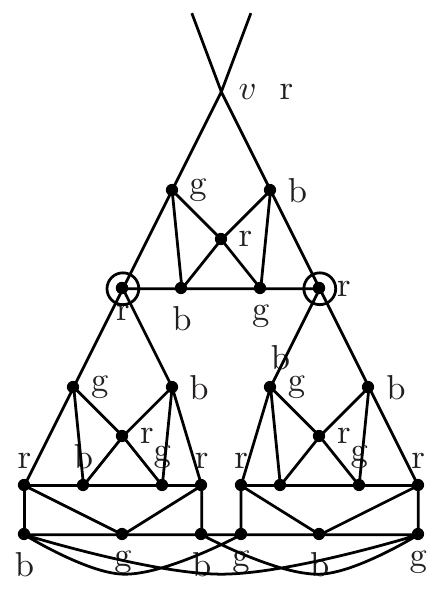}
  \caption{  }\label{fig:fig2}\vspace{-3mm}
\end{figure}

It is easy to see that the ear is regular of degree
4, is 3-colorable (as shown in Figure \ref{fig:fig2}
  with colors {\tt r,g,b}), but not
  planar. Hanging the ear adds a constant number of vertices. Now
  we have to deal with the set $V_\nit{odd}$ of vertices of degree 1 or 3 (vertices of degree 0 can be
  ignored). By Euler's theorem, $V_\nit{odd}$ has an even
  cardinality. This makes it possible to pick up
disjoint pairs $\{v_1, v_2\}$ of elements of $V_\nit{odd}$,
leaving every vertex coupled to some other vertex.
 For each such pair, $\{v_1,
v_2\}$, add an extra vertex $v'$ connected to (only) $v_1$ and
$v_2$.  This trio is 3-colorable.

Now $v_1, v_2$ have degree 2 or 4. ~~From
those that become of degree 2, hang the ``ear" as before. In this
way, all the nodes become of degree 4. The number of added
vertices is polynomial in the size of the original graph. The
4-colorability of $G'$ follows from the 4-colorability of $G$
(every planar graph is 4-colorable) and the 4-colorability of the
hanging ears.}\\

Notice that graph $G$, due to its planarity, is 4-colorable. The
graph $G'$, is an extension of graph $G$ that may not be planar,
but preserves 4-Colorability.  We use the construction in Lemma
\ref{lem:3col4reg} as follows: Given any planar graph $G$ of
degree 4,  construct graph $G'$ as in the lemma, which is regular
of degree 4 and 4-colorable. Its 4-colorability is encoded as a
database problem with a fixed set of  first-order constraints.
Since $G'$ is 4-colorable, the database is consistent.
Furthermore, $G'$ uses all the 4 colors in the official table of
colors, as specified by the ICs. In the update part, deleting one
of the  colors leaves us with the problem of coloring $G'$ with
only three colors  (under an A-repair semantics only changes of
colors are allowed to restore consistency), which is possible iff
the original graph $G$ is 3-colorable. Deciding about the latter
problem is $\nit{NP}$-complete \cite{gjs-1976}. We obtain

\begin{theorem}\label{th:incr-set-del-ins} \em For
 ground atomic queries, first-order ICs, and
update sequences consisting of tuple deletions, incremental CQA
under the A-repair semantics is \textit{coNP}-hard. 
\end{theorem}

  \dproof{Theorem \ref{th:incr-set-del-ins}}{If the update operation $U$ is a
  $\nit{delete}$ of
  a database atom, we reduce to our problem
  3-Colorability of
  planar graphs $G$ with vertex degree at most 4, which is
  \textit{NP}-complete \cite{gjs-1976}. Given such a non-empty graph $G$, we construct
  graph $G'$ as in
  Lemma \ref{lem:3col4reg}, which is also 4-colorable (because $G$ is and the
ears too).

Let  $E(X, Y)$ be a database relation encoding the edges of the
graph, $\nit{Coloring}$ a 2-ary database relation storing a
coloring of the vertices, and $\nit{Colors}$ a unary relation
storing the four colors
  allowed. Notice that a 4-coloring of $G$ can be found in polynomial time
  \cite{rsst-96}. Then also a 4-coloring for $G'$ can be found in polynomial
time (a 4-coloring for the ears can be given once and for all). The ICs,
essentially denials and inclusion dependencies, are as
  follows:
  \begin{enumerate}
  \item Every node is colored:~ $\forall xy \exists z (E(x,y) \rightarrow
  \nit{Coloring}(x, z))$.
  \item
  Nodes have one color: $\forall x y_1 y_2 \neg(\nit{Coloring}(x, y_1) \land
  \nit{Coloring}(x, y_2) \land y_1 \not = y_2)$.

  \item
  Colors must be allowed:~
~$\forall xy (\nit{Coloring}(x,y) \rightarrow \nit{Colors}(y))$.

  \item
  Vertex degree is not less than 4:~
  $\forall x ( \exists y E(x,y) \rightarrow \exists y_1 y_2 y_3 y_4 (E(x,y_1)
  \land E(x,
  y_2) \land E(x, y_3) \land E(x, y_4) \land y_1 \not = y_2
  \land y_1 \not = y_3 \land y_1 \not = y_4 \land y_2 \not = y_3 \land
  y_2 \not = y_3 \land y_3 \not = y_4))$.

  \item
  Vertex degree is not bigger than 5:~
  $\forall x y_1 \cdots y_5 \neg (E(x, y_1) \land \cdots \land
  E(x,y_5) \land y_1 \not = y_2 \cdots \land y_4 \not =  y_5)$.

  \item
  Only vertices are colored:~
  $\forall xy\exists z(\nit{Coloring}(x,y) \rightarrow E(x,z))$.

  \item All colors are used:~
  $\forall x\exists z (\nit{Colors}(x) \rightarrow \nit{Coloring}(z,x))$.
  \item $E$ is symmetric:~ $\forall xy(E(x,y) \rightarrow E(y,x))$.
  \item Adjacent vertices have different colors:\\ $\forall
  xyuw \neg (E(x,y) \wedge \nit{Coloring}(x,u) \wedge
  \nit{Coloring}(y,w)
\wedge u = w)$.
  \end{enumerate}
  The initial database $D$ stores the graph  $G'$, together with its
  4-coloring (that does use all 4 colors). This is a consistent instance.

  For the incremental part, if the update $U$ is the deletion of a color,
  e.g. $\nit{delete}_\nit{Colors}(c)$, i.e. of
  tuple $(c)$ from $\nit{Colors}$,  the instance becomes inconsistent, because an
inadmissible color  is being used in the coloring. Since repairs
can be obtained by changing attribute values in existing tuples
only, the only possible
  repairs are the 3-colorings of
  $G'$ with the 3 remaining colors  (if such colorings exist), which are obtained
  by changing colors in the second attribute of $\nit{Coloring}$.
  If there are no colorings, there are no repairs.

The query $Q\!:~ \nit{Colors}(c)?$ is consistently true only in
case there is no 3-coloring of the original graph $G$, because it
is true in the empty set of repairs.}\\

To obtain this result it is good enough to use the sum as the
aggregation
 function and the weight function
 that assigns
$1$ to each change.  Clearly, this lower bound also applies to
update sequences containing any combination of ~$\nit{insert,
~delete, ~change}$.

\section{Conclusions} \label{sec:concl}
The dynamic scenario for consistent query answering that considers
possible updates on a database
 had not been considered before in
the literature. Doing incremental CQA on the basis of the original
database and the sequence of updates is an important and natural
problem. Developing algorithms that take into account previously
obtained consistent answers that are possible cached and the
updates at hand is a crucial problem for making CQA scale up for
real database applications. Much research is still needed in this
direction.

In this paper we have concentrated mostly on complexity bounds for
this problem under different semantics. When we started obtaining
results for incremental CQA under repairs that differ from the
original instance by a minimum number of tuples, i.e. C-repairs,
we realized that this semantics had not been sufficiently explored
in the literature in the static version of CQA, and that a full
comparison was not possible. In the first part of this paper we
studied the complexity of CQA for the C-repair semantics and
denial constraints. In doing so, we developed  graph-theoretic
techniques for polynomially reducing each of the certain and
possible (or cautious and brave) C-repair semantics for CQA to the
other. A similar result does not hold for the S-repair semantics,
conjunctive queries, and denial constraints: CQA (under the {\em
certain} semantics) is $\nit{coNP}$-complete \cite{cm-2005}, but
is in $\nit{PTIME}$ for the {\em possible} semantics.

The complexity of CQA in a P2P setting was studied in
\cite{GrecoScarcello}, including a form a cardinality-based
repairs. However, a different semantics is used, which makes it
difficult to compare results. Actually, in that setting it is
possible that repairs do not exist, whereas in our case, since
S-repairs always exist \cite{abc-1999}, also C-repairs exist. The
complexity result for CQA in \cite{GrecoScarcello}, that seems to
be shared by C- and S-repairs, is obtained on the basis of the
complexity of checking the existence of repairs (a problem that in
our case is trivial).

 The C-repair semantics can be generalized considering
weights on tuples. Under denial constraints, this means that it
may be more costly to remove certain tuples than others to restore
consistency. More precisely, database tuples $R(\bar{t})$ have
associated numerical costs $w(R(\bar{t}))$, that become part of
the input for the CQA decision problem.  Now, the partial order
between instances is given by $D_1 \preceq_{D,wC} D_2$ iff $|D
\triangle D_1|_w \leq |D \triangle D_2|_w$, where, for a set of
database tuples $S$,  $|S|_w$ is the sum of the weights of the
elements of $S$. It can be proved that CQA for ground atomic
queries wrt denial constraints under this semantics belongs to
$P^\nit{N\!P}$ \cite[proposition 5]{corr}.

Furthermore, it possible to reduce CQA under the C-repair
semantics to CQA under {\em least-squares} A-repairs semantics
 that minimizes the sum of the quadratic
differences between numerical values \cite{bbfl-2005}, which is a
particular case of the general semantics studied in Section
\ref{sec:attrib}.

\begin{theorem} \label{theo:redCA} \em
Given a database schema $\cal D$, a set $\IC$ of denial
constraints in $L({\cal D})$, and a ground atomic query $Q \in
L({\cal D})$, there are a schema ${\cal D}'$ with some fixable
numerical attributes, a set $\IC'$ of ICs in $L({\cal D}')$, and a
query $Q' \in L({\cal D}')$, such that: For every database $D$
over ${\cal D}$, there is a database $D'$ over ${\cal D}'$ that
can be computed from $D$ in {\em LOGSPACE} (in data) for which it
holds: $Q$ is consistently true wrt $\IC$ in $D$ under the
C-repairs semantics iff $Q'$ is consistently true wrt to $\IC'$ in
$D'$ under the least-squares A-repair semantics. 
\end{theorem}

\dproof{Theorem \ref{theo:redCA}}{Given a schema $\cal R$
with relations $R_1, \ldots, R_m$ for CQA under the C-repair semantics,
expand each relation $R_i$ to $\bar{R}_i$ that has an extra attribute $E_i$
that takes numerical values $0$ or $1$, and is the only fixable attribute
for $\bar{R}_i$. Transform each denial of the form
$\forall \bar{x} \neg( \cdots R_i(\bar{t}) \cdots)$
into the denial
$\forall \bar{x} \cdots \forall e_i \cdots \neg(\cdots \bar{R}_i(\bar{t},e_i) \wedge e_i = 1 \cdots)$.

An atomic query $R_i(\bar{t})$ for CQA under the C-repair semantics is
transformed into $R_i(\bar{t},1)$, which is answered under the least-squares
A-repair semantics.

An instance $\bar{D}$ is created from an instance $D$ for $\cal R$, by
inserting $\bar{R}_i(\bar{c},1)$ into $\bar{D}$ when $R_i(\bar{c}) \in D$.
}\\

This result also applies to other numerical A-repair semantics as
discussed in \cite{bbfl-2005}, and is about data complexity. For
fixed ${\cal D}, \IC, Q, D$, also fixed ${\cal D}', \IC', Q'$ can
be obtained in {\em LOGSPACE} from ${\cal D}, \IC, Q$. Theorem
\ref{theo:redCA}, together with Theorem \ref{teo:hardness}, allows
us to obtain a simple proof of the $P^\nit{NP(log~n)}$-hardness of
the least-squares repair semantics. In \cite{bbfl-2005},
$P^\nit{NP}$-hardness is obtained for the latter as a better lower
bound, but the proof is more complex. This theorem can be extended
to the weighted C-repair semantics if integer numerical weights
are used.

Our results show that the incremental complexity is lower than the
static one in several useful cases, but sometimes the complexity
cannot be lowered. It is a subject of ongoing work the development
of concrete and explicit algorithms for incremental CQA.

We obtained the first results about fixed parameter tractability
for incremental CQA, where the input, for a fixed database schema,
can be seen as formed by the original database and the update
sequence, whose length is the relevant parameter. This problem
requires additional investigation. In particular, the
parameterized complexity of incremental CQA under the S- and
A-repair semantics has to be investigated, and a more complete
picture still has to emerge.

It would be interesting to examine the area of CQA in general from
the point of view of parameterized complexity, including the
static case. Natural candidates to be a parameter in the classic,
static setting could be: (a) the number of inconsistencies in the
database, (b) the degree of inconsistency, i.e. the maximum number
of violations per database tuple, (c) complexity of inconsistency,
i.e. the length of the longest path in the conflict graph or
hypergraph. These parameters may be practically significant, since
in many applications, like census application \cite{bbfl-2005},
inconsistencies are ``local".

We considered a version of incremental CQA that assumes that the
database is already consistent before updates are executed, a
situation that could have been achieved because no previous
updates violated the given semantic constraints or a repaired
version was chosen before the new updates were executed.

We are currently investigating the dynamic case of CQA in the
frameworks of  \emph{dynamic complexity} \cite{imm-99,Schwentick}
and \emph{incremental complexity} as introduced in
\cite{msvt-1994}. In this case we start with a database $D$ that
is not necessarily consistent on which a sequence of basic update
operations $U_1, U_2, ..., U_m$ is executed. A clever algorithm
for CQA  may create or update intermediate data structures at each
atomic update step, to help obtain answers at subsequent steps. We
are interested in the complexity of CQA after a sequence of
updates, when the data structures created by the query answering
algorithm at previous states are themselves updatable and
accessible.

\vspace{2mm} \noindent {\bf Acknowledgments:}~~ Research supported
by NSERC, and EU projects:  Knowledge Web, Interop and Tones. ~L.
Bertossi is Faculty Fellow of  IBM Center for Advanced Studies
(Toronto Lab.). L. Bertossi  appreciates the hospitality and
support of Enrico Franconi and the KRDB group in Bolzano. We are
grateful to Jan Chomicki, J\"org Flum, and anonymous referees for
many useful comments.

\bibliographystyle{plain}

\ignore{
\newpage

\section{Appendix: Basic Notions of Complexity Theory}

We briefly recall some of the complexity classes used in this
paper. $\nit{FP}$ is the class of functional problems  that are
solvable in polynomial time. $P^\nit{NP}$  (or $\Delta_2^P$) is
the class of decision problems solvable in polynomial time by a
machine that makes calls to an $\nit{NP}$ oracle.
$P^\nit{NP(log(n))}$ is similarly defined, but the number of calls
is logarithmic. It is not known if $P^\nit{NP(log(n))}$ is
strictly contained in $P^\nit{NP}$. The functional class
$\nit{FP}^\nit{NP(log(n))}$ is similarly defined. The class
$\Delta^P_3\!(\nit{log(n)})$ contains decision problems that can
be solved by a polynomial time machine that makes a logarithmic
number of calls to an oracle in $\Sigma^P_2$. For more details we
refer to \cite{papadimitriou94}; and to \cite{flum} for
parameterized complexity.

\section{Appendix: Proofs}\label{sec:proofs}

\ignore{
\defproof{Lemma \ref{lem:graphExt}}{(a) We consider the three cases for membership
of $v$ to maximum
independent sets in $G$. Let $m$ be the cardinality of a maximum
independent set in $G$. We establish now the first bi-conditional.
The second
bi-conditional follows directly from the analysis for the first one. \\
(a) Assume that $v$ belongs to a maximum independent set $I$ of
$G$. In this case, $v'$ can be added to $I$ obtaining an
independent set of $G'$. In this case $|I \cup \{v'\}| \geq m+1$.

Assume that $v$ does not belong a some maximum independent set
$I'$ of $G'$. If $v \notin I'$, then some of its neighbors belong
to $I'$, and then, $v' \notin I'$. In consequence, $I'$ is also a
maximum independent set of $G$. Then, $|I'| = m$. But this is not
possible, because the size of independent set of $I'$ is at least
$m +1$.\\
(b) Assume that $v$ does not belong to any maximum independent
sets of $G$. Then, some of it neighbors can be found in every
maximum independent set of $G$, and none of them can be extended
 with $v'$ to become an independent set of $G'$.

 So, all the maximum independent set of $G$ are maximum independent sets of
 $G'$ of size $m$.

 Assume, that $v$ belongs to all maximum independent sets of $G'$.
  Then none of the neighbors of $v$ can be found in
 independent sets of $G$, and then
$v'$ can be found in all the maximum independent sets of
 $G'$. Since the maximum independent sets of $G'$ have at least
 cardinality $m$, it must hold that the maximum independent sets
 of $G'$ have cardinality at least $m+1$. Then the deleting $v'$
 from all the maximum independent sets of $G'$ will give us
 independent sets of $G$ of size at least $m$, i.e. maximum
 independent sets of $G$. To all of them $v$ belongs. A
 contradiction.\\
 (b) (sketch) Hang a rhombus from $v$, i.e.
 add three other vertices, two of them connected to $v$, and the
 third one, connected to the two previous ones. Then,
 reason by cases as in the proof of part (a).}

\vspace{-3mm}
\defproof{Lemma \ref{lem:dec-graphNPlogn}}{For the first claim, given a
graph $G$ and a vertex $v$, build in polynomial time the graph
$G'$ as in Lemma \ref{lem:graphExt}(a). It holds that $v$ belongs
to some maximum independent set of $G$ iff $v$ belongs to every
maximum independent set of $G'$. Now, $v$ belongs to every maximum
independent set of $G'$ iff $|\mbox{maximum independent set}$
$\mbox{in } G'| - |\mbox{maximum independent set in } G| = 1$.

Since computing the maximum cardinality of a clique can be done in
time $\nit{FP}^\nit{NP(log(n))}$ \cite{krentel} (see also
\cite[theorem 17.6]{papadimitriou94}), computing the maximum
cardinality of an independent set can be done in the same time
(just consider the complement graph). In consequence, in order to
decide about $v$ and $G$, we can compute the cardinalities of the
maximum independent set for $G$ and $G'$ in 2 times
$\nit{FP}^\nit{NP(log(n))}$, and next compute their difference. It
total, we can perform the whole computation in
$\nit{FP}^\nit{NP(log(n))}$. In consequence, by definition of
class $\nit{FP}^\nit{NP(log(n))}$, we can decide by means of a
polynomial time machine that makes $O(\nit{log}(n))$ calls to an
$\nit{NP}$ oracle, i.e. the decision is made in time
$P^\nit{NP(log(n))}$. The same proof works for the second claim.
It can also be obtained from the first claim and Lemma
\ref{lem:graphExt}(b).}
}

\ignore{\vspace{-3mm}
\defproof{Theorem \ref{prop:memb}}{Construct the conflict
graph for the instance wrt the FDs. An atomic ground query is
consistently true if the corresponding vertex in the conflict
graph belongs to all the maximum independent sets.  Then use
Lemma \ref{lem:dec-graphNPlogn}.}
}

\ignore{
\defproof{Lemma \ref{lemma:allRed}}{We prove: (1) $\Rightarrow$ (2) $\Rightarrow$ (3) $\Rightarrow$
(4). That (4) $\Rightarrow$ (1) was established in Lemma
\ref{lem:graphExt}(a).\\
(1) $\Rightarrow$ (2):~ Given a graph $G$ and a vertex $v$, extend
$G$ to  a graph $G'$ by adding  new vertices $s, s'$ with $s$
connected to $v$ and $s'$ connected to $s$.
If $v$ belongs to every MIS of $G$, then $s$ does not belong to any MIS of $G'$.\\
If $v$ belongs to one MIS of $G$, but  not to the others, then $s$ belongs to some MIS of $G'$
 and does not belong any other MIS of $G'$.
If $v$ does not belong to any MIS of $G$, then $s$  belongs to one
MIS of $G'$ but not to the others. Thus, $v$ belong to every MIS
of $G$ if and only if $s$ does not belong to any MIS of $G'$.
\\
(2) $\Rightarrow$ (3):~ Given $G, v$, extend $G$ to $G'$ by adding
a vertex $s$ and connecting it to $v$. If $v$ belongs to every MIS
of $G$, then $s$ belongs to every MIS of $G'$. If $v$ belongs to
one MIS of $G$ but and not to the others, then $s$ belongs to
every MIS of $G'$. If $v$ does not belong to any MIS of $G$, then
either $s$ does not belong to any MIS of $G'$ or there is a MIS of
$G'$  to which $s$ does not belong (depending on $G$).
\\
(3) $\Rightarrow$ (4):~ Given $G, v$, extend $G$ to $G'$ by adding
vertices $s_1, s_2, s_3, s$, and the edges $\{s_1,v\},
\{s_2,s_1\}, \{s_3,s_1\}, \{s,s_2\}, \{s,s_3\}$. If $v$ belongs to
every MIS of $G$, then $s$ does not belong to any MIS of $G'$. If
$v$ belongs to one MIS of $G$ but not to the others, then $s$
belongs to one MIS of $G'$ but not  to the others. If $v$ does not
belong to any MIS of $G$, then $s$ belongs to one MIS of $G'$ but
not to the others. }

}

\ignore{
\vspace{-3mm}
\defproof{Theorem \ref{prop:membDCs}}{We use the
conflict hypergraph. The problem of determining the maximum clique
size for hypergraphs is in $\nit{FP}^{\nit{NP}(\nit{log}(n)}$ by
the same argument as for conflict graphs: Deciding  if the size of
maximum clique is greater than $k$ is in $\nit{NP}$. So, by asking
a logarithmic number of $\nit{NP}$ queries, we can determine the
size of maximum clique.

The membership to $P^{\nit{NP}(\nit{log}(n))}$ of CQA for the
C-repair semantics
 still holds for conjunctive queries
without existential variables. In fact, given an inconsistent
database $D$, a query $Q$, and a ground tuple $t$, we check if $t$
is consistent answer to $Q$ from $D$ as follows: Check if $t$ is
an ordinary answer to $Q$ in $D$ (without considering the
constraints). If not, the answer is {\em no}.

Otherwise, let $t_1, \ldots, t_k$ be the database tuples which are
answers to $Q$ in $D$ and produce $t$ as an answer. Since $Q$ does
not contain existential variables, only one such set exists.
Compute the size of a maximum independent set for the graph
representation of $D$, say $m_0$. Compute the size of a maximum
independent set for the graph representation of $D \smallsetminus
\{t_1\}$, say $m_1$. If $m_1 = m_0$, then there exist a maximum
independent set of $D$ that does not contain $t_1$. So, there
exists a minimum repair that does not satisfy that $t$ is an
answer to $Q$. If $m_1 < m_0$, repeat this procedure for all
tuples in $t_1, \ldots, t_k$. Thus, we have to pose $k$ queries
(that is determined only by the size of the query) to an
$\nit{FP}^{\nit{NP(log}(n))}$ oracle.

In consequence,  CQA for conjunctive queries without existential
variables is in $\nit{P}^{\nit{NP(log}(n))}$.}
}

\ignore{
\vspace{-3mm}
 \defproof{Lemma \ref{lem:everyGraph}}{Consider a graph $G = \langle V,
E\rangle$, and assume the vertices of $G$ are uniquely labelled.
Consider the database schema with two relations, $\nit{Vertex}(v)$
and $\nit{Edges}(v_1, v_2, e)$, and the denial constraint $\forall
v_1 v_2 \neg(\nit{Vertex}(v_1) \land \nit{Vertex}(v_2) \land
\nit{Edges}(v_1, v_2,e))$. $\nit{Vertex}$ stores the vertices of
$G$. For each edge $\{v_1,v_2\}$ in $G$, $\nit{Edges}$ contains
$n$ tuples of the form $(v_1, v_2,i)$, where $n$ is the number of
vertices in $G$. All the values in the third attribute of
$\nit{Edges}$ are different, say from $1$ to $n |E|$. The size of
the database instance obtained trough this padding of $G$ is still
polynomial in size.

This instance is highly inconsistent, and  its C-repairs are all
obtained by deleting vertices, i.e. elements of $\nit{Vertex}$
alone. In fact, an instance such that all tuples but one in
$\nit{Vertex}$ are deleted, but all tuples in $\nit{Edges}$ are
preserved is a consistent instance. In this case, $n-1$ tuples are
deleted. If we try to achieve a repair by deleting tuples from
$\nit{Edges}$, say $(v_1,v_2,i)$, then in every repair of that
kind all the $n$ tuples of the form $(v_1, v_2,j)$ have to be
deleted as well. This would not be a minimal cardinality repair.

Assume that $I$ is a maximum cardinality independent set of $G$.
The deletion of all tuples $(v)$ from $\nit{Vertex}$, where $v$
does not belong to $I$, is a C-repair. Now, assume that $D$ is a
repair. As we know, only tuples from $\nit{Vertex}$ may be
deleted. Since, in order to satisfy the constraint, no two
vertices in the graph that belong to $D$ are adjacent, the
vertices remaining in $\nit{Vertex}$ form an independent set in
$G$.

In general, the number of deleted tuples is equal to $n-|I|$,
where $I$ is an independent set represented by a repair. So each
minimal cardinality repair corresponds to a maximum independent
set and vice-versa.}

\vspace{-3mm}
\defproof{Theorem \ref{cor:size}}{This follows from Lemma
\ref{lem:everyGraph}, the fact
that C-repairs correspond to maximum cliques in the complement of
the conflict graph \cite{abchrs-2003}, and the
$P^\nit{NP(log(n))}$-completeness of determining the size of a
maximum clique \cite{krentel}.}
}

\ignore{
\vspace{-3mm}
\defproof{Lemma \ref{lem:block}}{
The new graph $G'$ consists of two copies of $G$, say $G_1, G_2$,
two additional graphs, $I_k, I_{k+1}$, and two extra vertices
$t,b$. Subgraph $I_k$ consists of $k$ mutually disconnected
vertices; subgraph $I_{k+1}$ consists of $k+1$ mutually
disconnected connected vertexes. Each vertex of $G_1$ is adjacent
to each vertex of $I_k$, and each vertex of $G_2$ is adjacent to
each vertex of $I_{k+1}$. Each vertex of $I_k$ is adjacent to $t$,
and each vertex of $I_{k+1}$ is adjacent to $b$.  Finally, $t, b$
are connected by an edge (c.f. Figure 1).

 We claim that vertex $t$ belongs to all maximum independent sets
of $G'$ iff the cardinality of maximum independent set of $G$ is
equal to $k$. To prove this claim, we consider a few, but
representative possible cases. With $I(G)$ we denote an arbitrary
maximum independent set of $G$.
\begin{itemize}
 \item [1.] $|I(G)| < k - 1$: The maximum independent set of $G'$ is
 $I_k \cup I_{k+1}$; with cardinality
 $2k + 1$.
 \item [2.] $|I(G)| = k - 1$: The maximum independent sets of $G'$
 are (a) $I(G_1) \cup I_{k+1} \cup
 \{t\}$, and
 (b) $I_k \cup I_{k+1}$, with cardinality $2k+1$.
 \item [3.] $|I(G)| = k$: The maximum independent set of $G'$ is
 $I_{k+1} \cup I(G_1) \cup \{t\}$, with cardinality $2k+2$.
 \item [4.] $|I(G)| = k + 1$: The maximum independent sets of $G'$ are
 (a) $G_1 \cup G_2 \cup \{t\}$,  (b) $G_1 \cup G_2
 \cup \{b\}$, (c) $G_1 \cup I_{k+1} \cup \{t\}$; with cardinality $2k+3$.
 \item [5.] $|I(G)| > k + 1$: The maximum independent sets of $G'$ are
 (a) $G_1 \cup G_2 \cup \{t\}$, (b) $G_1 \cup G_2 \cup
\{ b\}$;
 with cardinality $2|I| +1$.
\end{itemize}
Only in case $|I(G)| = k$, $t$ belongs to all maximum independent
sets.}

\vspace{-3mm}
\defproof{Lemma \ref{prop:allsets}}{
By reduction from the following
$P^{\nit{NP}(\nit{log}(n))}$-complete decision problem
\cite[theorem 3.5]{krentel}: Given a graph $G$ and an integer $k$,
is the size of a maximum clique in $G$ equivalent to $0~
\nit{mod}~ k$?

Assume graph $G$ has $n$ vertices. We can also assume that $k$ is
not bigger than $n$. Now, we pass to the graph $G'$ that is the
complement of $G$: It has the same vertices as $G$, with every two
distinct vertices being adjacent in $G'$ iff they are not adjacent
in $G$. A maximum independent set of $G'$ is a maximum clique of
$G$ and vice-versa. So, the cardinality of a maximum independent
set of $G'$ is the size of a maximum clique of $G$.

Next, we take advantage of the construction in Lemma
\ref{lem:block} (c.f. Figure 1): For each $m \in \{k, 2k, \cdots,
\lfloor \frac{n}{k} \times k\rfloor \}$, construct the block graph
$B_m(G', t_m)$. (There are $[n/k]$ possible solutions to the
equation $x \equiv 0 ~\nit{mod}~k$.) All these graphs are
disconnected from each other. Next, create a new vertex $t_g$ and
connect it to the vertices $t_m$  of the blocks $B_m(G',t_m)$. It
is easy to check that the resulting graph, say $\overline{G}$, has
its size bounded above by $O(n^4)$.

It holds that vertex $t_g$ does not belong to every maximum
independent set of $\overline{G}$ iff the size of maximum
independent set of $G$ is equivalent to $0\ \nit{mod}\ k$. So, we
have a reduction to the complement of our problem, but the class
$P^{\nit{NP(log}(n))}$ is closed under complement.

In fact, if the size of maximum independent set of $G$ is not
equivalent to $0~ \nit{mod}~k$, then for every block $B$ in
$\overline{G}$, there exists a maximum independent set $I_B$ of
the block $B$ such that $t_B \notin I_B$ ($t_B$ is the top node of
block $B$). The maximum independent set of $\overline{G}$ is
$\{t_g\} \cup \bigcup_B I_B$ (because there are no edges between
blocks and between $t_g$ and other vertices besides $t_B$).
Consider any independent set $I$ of $\overline{G}$ that does not
contain $t_g$. The size of the projection of $I$ on any block is
not greater than the size of the maximum independent set of the
block; so $|I| \leq |\bigcup_B I_B|$. So, $t_g$ belongs to every
maximum independent set of $\overline{G}$.

Now, if the size of a maximum independent  set  of $G$ is
equivalent to $0\ \nit{mod}\ k$, then there exists one block
$B_{\!o}$ such that $t_{B_o}$ belongs to every maximum independent
set $I_{B_o}$ of $B_{\!o}$, while for all other blocks $B$ there
exists $I_B$ such that $t_B \notin I_B$. Consider a maximum
independent set $I_t$ of $\overline{G}$ that contains $t_G$.

Every maximum independent set of $\overline{G}$ that contains
$t_g$ is of the form $\{t_g\}$ union of maximum independent sets
from the blocks $B$ other than $B_o$ that do not contain their
corresponding $t_B$ union any maximum independent set of $B_{\!o}
\smallsetminus \{t_{B_o}\}$. The size of such a set is $s = 1 +
\sum_{B \neq B_o} |I(B)| + (|I_{B_o}| -1)$. A maximum independent
set $I$ that does not contain $t_g$, is the union of maximum
independent sets $I_B$ of all the blocks $B$ of $\overline{G}$,
and its size is equal to $\sum_B |I_B|$, i.e. $s$. Then, there
exists a maximum independent set that does not contain $t_g$. }
}

\ignore{
\vspace{-3mm}
\defproof{Theorem \ref{teo:hardness}}{Membership follows from
Theorem \ref{prop:membDCs}. Now we prove hardness.
 For a graph $G$ and integer $k$, we
construct a database $D$, such that the consistent answer to a
ground atomic query $Q$ can be used to decide if the size of a
maximum clique of $G$ is equivalent to $0~ \nit{mod}~ k$ (c.f.
proof of Lemma \ref{prop:allsets}). Construct the graph
$\overline{G}$ as in Lemma \ref{prop:allsets}. Encode graph
$\overline{G}$ as a database inconsistency problem, introducing a
unary relation $V$ (for vertices) and $E$ (3-ary), where $E$
corresponds to the edge relation in $\overline{G}$ plus a third
padding attribute to make changing it more costly. For each vertex
$v \in \overline{G}$, there is a tuple $(v)$ in $V$.

We also introduce the denial constraint: $\forall v_1 \forall v_2
\neg(V(v_1) \wedge V(v_2) \wedge E(v_1, v_2,\!\_))$ (an underscore
means any variable implicitly universally quantified).
 For each edge
$\{v_1, v_2\} \in \overline{G}$, create $n$ different versions
$(v_1, v_2,p)$ in $E$, as in the proof of Lemma
\ref{lem:everyGraph}. The effect of fixing the database wrt the
given denial constraint may be the removal of tuples representing
vertices or/and the removal of  tuples representing edges. We want
to forbid the latter alternative because those repairs do not
represent maximum independent set; and this is achieved by making
them more expensive than vertex removal through the padding
process.

The consistent answer to the query $V(t_g)$ is {\em no}, i.e. not
true in all repairs, iff $t_g$ does not belong to all maximum
independent sets of $\overline{G}$ iff the size of a maximum
independent set of $G'$ is equivalent to $0~ \nit{mod}~ k$ iff the
size of a maximum clique of $G$ is equivalent to $0~ \nit{mod}~
k$.}
}

\ignore{
\vspace{-3mm}
\defproof{Proposition \ref{theo:weightedpro}}{Membership is proved as for the
C-repair semantics (c.f.  Theorem \ref{prop:membDCs}): Repairs
correspond to maximum weighted independent sets of the associated
hypergraph $G$. The weight of a maximum weighted independent set
can be found in $P^\nit{NP}$ (as for  independent set, but
$\nit{log}(O(2^n)) = \nit{poly}(n)$ oracle calls are required). To
check if a vertex $v$ belongs to all maximum weighted independent
sets, it is good enough to compute weights of maximum independent
sets for $G$ and $G \setminus \{v\}$.}
}

\ignore{
\vspace{-3mm}
\defproof{Theorem \ref{prop:incr-dens}}{
For denial constraints tuple deletions do not introduce any
violations, so we consider a sequence $U$ consisting of tuple
insertion and updates.

 Assume that $k$ of the $m$ inserted
tuples violate ICs, perhaps together with some tuples already in
$D$. If we delete $k$ violating tuples, then we get a consistent
database $D'$;  so a minimal repair is at a distance less than or
equal to $k$ from $D$. To find all minimal repairs it is good
enough to check no more than $N = {\small \left(
\begin{array}{c} n + m \\ 1 \end{array} \right) + \left(
\begin{array}{c} n + m\\ 2
\end{array} \right) + \cdots\ + \left( \begin{array}{c} n + m\\ k
\end{array} \right)}$
 repairs, where $|D| = n$. If $m$ is small, say less than $c
\cdot n$, then $N < k {\small \left( \begin{array}{c} n + m\\
k\end{array}\right)}  \leq m {\small \left( \begin{array}{c} n\\
m\end{array}\right)}^{\!\!m} <  mn^m$. Thus, the incremental
complexity of the CQA is polynomial wrt $n$.

In case $U$ contains change updates, the proof is essentially the
same, but the role of $m$ is taken by $m \cdot a$, where $a$ is
the maximum arity of the relations involved. This is because we
have to consider possible changes in different attributes.}
}

\ignore{
\vspace{-3mm}
\defproof{Theorem \ref{prop:fpt}}{
First, it is known that the problem of, given a graph $G$ and a
number $k$, determining if there exists a vertex cover of size
less than or equal to $k$ is in FPT \cite{downfel}. We will use
this problem to solve ours.

Now, let us assume that we have a consistent database $D$ of size
$n$, and we update it inserting $k$ new tuples, obtaining an
inconsistent database $D'$ with conflict graph $G$. The size of
$G$ is $O(n)$ by our assumption on the size of $m$ in comparison
with $n$. Every C-repair of $D'$ is a maximum independent set of
$G$, and can be obtained by deleting from $G$ a minimum vertex
cover, because the problems are complementary. So, a minimum
vertex cover corresponds to the vertices that are to be deleted to
obtain a repair.

Since the original database $D$ is consistent, the vertices of $G$
corresponding to database tuples in $D$ are all disconnected from
each other. In consequence, edges may appear only by the update
sequence, namely between the $m$ new tuples or between them and
the elements of $D$. Then, we know that there is a vertex cover
for $G$ of size $m$. However, we  do not know if it is minimum.

In order to find the size of a minimum vertex cover of $G$, we may
start doing binary search from $m$, applying an FPT algorithm for
vertex cover.  Each check for vertex cover, say for value $m_i$,
can be done in $O(1.2852^m_i + m_i \cdot n)$ \cite{chen}. Then
$\nit{log}(m)$ checks take time $O(\nit{log}(m) \cdot (1.2852^m +
m \cdot n)) \leq O(f(m) \cdot n)$, with $f$ an exponential
function in $m$. So, it is in $\nit{FPT}$ obtaining the size of a
minimum vertex cover for $G$, which gives us the minimum number of
tuples to remove to restore consistency.

Now, for CQA we want to check if a vertex $R(\bar{t})$ belongs to
all maximum independent sets of $G$, which happens if it does not
belong to any minimum vertex covers. This can be determined by
checking the size of minimum vertex cover for $G'$ and $G'
\smallsetminus \{R(\bar{t})\}$. If they are the same, then
$R(\bar{t})$ belongs to all maximum independent sets and the
consistent answer to the query $R(\bar{t})$ is $\nit{yes}$.}
}

\ignore{
\vspace{-3mm}
\defproof{Proposition \ref{lem:param-compl-incr}}{
 By uniform
reduction from the $\nit{MONOTONE~W[1]}$-hard problem
\cite{df-1995} \emph{WEIGHTED MONOTONE 3CNF SAT}, which is defined
as follows: Given a 3CNF monotone circuit $C$ and an integer $k$,
is it possible to make exactly $k$ of the inputs $1$ and obtain
output $1$ for $C$?

The database schema consists of relations $\nit{Clause}(C, V_1,
V_2, V_3, p)$, $\nit{Var}(V)$, $\nit{Cond}(X,Y)$, where $p$ and
$Y$ are dummy variables intended to create many copies of a tuple,
to forbid the deletion of those tuples by making the potential
repair too costly.  The integrity constraint is~ $\forall C V_1
V_2 V_3 p y \neg (Clause(C, V_1, V_2, V_3, p) \land Var(V_1) \land
Var(V_2) \land Var(V_3) \land \nit{Cond}(1,y))$. Given a monotone
3CNF formula $\Psi = \psi_1 \land \psi_2 \land \cdots \land
\psi_m$ and a parameter $k$, for each clause $\psi_i = (x_{i_1}
\lor x_{i_2} \lor x_{i_3})$, where the $x_{i_j}$ are atoms,
 store in $\nit{Clause}$ $n$ copies of the form $(i, x_{i_1}, x_{i_2},
 x_{i_3},p)$
(replace variable by any new constant if a clause has less than
three variables). For each variable $x$ in $\Psi$, store $x$ in
$Var$. $C$ is initially empty. The resulting database is
consistent.

Now, on the update part, insert $(1, i)$ into $\nit{Cond}$, $i =
1, \ldots k$. Then there exists an assignment with weight less
than $k$ iff $\nit{Cond}(1,1)$ is $\nit{false}$ in every repair.

Since we have to determine if there exists a satisfying assignment
with weight  exactly $k$, it is good enough to ask a query to two
databases, built as before, but for both $k$ and $k+1$, which is
compatible with the definition of \emph{parametric reduction},
that allows to use of a constant number of instances. In our case,
since we have that: (a) if weight $< k$, then consistent answer is
{\em yes}, (b) if weight is equal to $k$, then the consistent
answer is \emph{false} (i.e. false in all repairs), and (c) if
weight $> k$, the consistent answer is \emph{false}. So, we
construct two instances, for $k$ and $k+1$. The weight is equal to
$k$ iff the consistent answer for the first instance is
\emph{false} and for the second one it is {\em yes}. }
}

\ignore{
\vspace{-3mm}
\defproof{Theorem \ref{lem:incr-icd-tds-msd}}{By reduction from static
  CQA for (existentially quantified) conjunctive queries and denial ICs under
  minimal set semantics, which is \textit{coNP}-hard
  \cite{cm-2005}. Consider an instance for this problem consisting of a database
  $D$, a set of denial ICs $\IC$, and a query $Q$.

  For every denial $\nit{ic} \in \IC$, pick up a relation $R^\nit{ic}$ in it and
  expand it to a relation
  $\overline{R^\nit{ic}}$ with an extra attribute $\nit{Control}$.
  Also add a new, one attribute relation
  $\nit{Controler}(A)$. Next, transform each integrity constraint
$\nit{ic}\!: ~
  \forall \bar{x} \neg (P(\bar{x}) \land \cdots \wedge R^\nit{ic}(\bar{x}) \land
  \cdots \wedge \gamma)$ into
  $\nit{ic}^{\prime}\!: ~\forall \bar{x}\forall \nit{contr} \neg (P(\bar{x}) \land
  \cdots \wedge
  \overline{R^\nit{ic}}(\bar{x},
  \nit{contr}) \land \nit{Controler(contr)} \land \gamma)$. We
  obtain a set $\IC'$ of denial constraints. The original database
  $D$ is extended to a database $\overline{D}$ with the new relation
  $\nit{Controler}$, which is initially empty, and the relations $\overline{R^\nit{ic}}$,
  whose extra attributes $\nit{Contr}$ initially take all
the value $1$. Due to the extension of $\nit{Controler}$, $\IC'$
is satisfied.

  Now in the incremental context, we consider the inconsistent instance $\overline{D}'$
  obtained via the update $\nit{insert}(\nit{Controler}(1))$ on $\overline{D}$.
The S-repairs  of $\overline{D}'$ wrt $\IC'$ are: (a)
$\overline{D}$ and (b) all the S-repairs
  of $\overline{D}$ (plus the tuple $\nit{Controler}(1)$ in each of them), which are in one-to-one
  correspondence with the S-repairs of $D$ wrt $\IC$.
Now, for a conjunctive query $Q$ in the language of $D$, produce
the conjunctive query
  $Q'\!: \exists \cdots y_\nit{ic} \cdots Q\frac{\cdots R^\nit{ic}(\bar{x}) \cdots}{\vspace*{1mm}\cdots
  \overline{R^\nit{ic}}(\bar{x},y_\nit{ic}) \cdots}$ in the
  language of $\overline{D}$,\footnote{$E\frac{E_1}{E_2}$ means the expression obtained by replacing
  in expression $E$ the subexpression $E_1$ by expression $E_2$.} where each atom
  $R^\nit{ic}(\bar{x}))$ in $Q$ is replaced by $\exists y_\nit{ic}
  \overline{R^\nit{ic}}(\bar{x},y_\nit{ic})$.

  Notice that all the repairs in (b) are essentially contained
 in $\overline{D}$, except for the tuple $\nit{Controler}(1)$, whose
  predicate does not appear in the queries. This is because denial
  constraints are obtained by tuple deletions. In consequence, any
  answer to the conjunctive (and then monotone) query in a repair
  in (b) is also an answer in the repair in (a). In consequence,
  the repair $\overline{D}$ does not contribute with any new
  consistent answers, neither invalidates any answers obtained by
  the repairs in (b). So, it holds $\nit{Cqa}(Q,D,\IC) =
  \nit{Cqa}(Q',\overline{D}',\IC')$.}
}

\ignore{
\vspace{-3mm}
  \defproof{Theorem \ref{theo:wmccs}}{ We provide a $\nit{LOGSPACE}$-reduction
from the following problem \cite[theorem 3.4]{krentel}: Given a
Boolean formula $\psi(X_1, \cdots, X_n)$ in 3CNF, decide if the
last variable $X_n$ is equal to $1$ in the lexicographically
maximum satisfying assignment (the answer is ${\it No}$ if $\psi$
is not satisfiable).

Create a database schema with relations:~ $\nit{Clause(id,
Var}_1,$ $ \nit{Val}_1, \nit{Var}_2, \nit{Val}_2,$ $\nit{Var}_3,
\nit{Val}_3)$, $\nit{Var(var, val)}$, $\nit{Dummy}(x)$, with
denial constraints:\\
$\forall var, val  \neg (Var(var, val) \land val \not = 0  \land
val \not = 1)$,\\
$\forall id, v_1, x_1, v_2, x_2, v_3, x_3 \neg (Cl(id, v_1, x_1,
v_2, x_2, v_3, x_3)
 \land Var(\_, v_1, x'_1)  \land Var(\_, v_2, x'_2)$
 $\land~ Var(\_, v_3, x'_3) ~\land~
 x_1 \not = x'_1  \land x_2 \not = x'_2 \land x_3 \not =  x'_3
\land
 \nit{Dummy}(1))$.

\noindent The last denial can be replaced by 8 denial constraints
without inequalities considering all the combination of values for
$x_1, x_2, x_3$ in $\{0,1\}$.

Assume now that $C_1, \ldots, C_m$ are the clauses in $\psi$. For
each propositional variable $X_i$ store in table $\nit{Var}$ the
tuple $(X_i, 0)$, with weight $1$, and $(X_i,1)$ with weight
$2^{n-i}$. Store tuple $1$ in $Dummy$ with weight $2^n\times 2$.
For each clause $C_i = l_{i_1} \lor l_{i_2} \lor l_{i_3}$, store
in $\nit{Clause}$  the tuple $(C_i, X_{i_1}, \tilde{l}_{i_1},
X_{i_2}, \tilde{l}_{i_2}, X_{i_3}, \tilde{l}_{i_3})$, where
$\tilde{l}_{i_j}$ is equal to $1$ in case of positive occurrence
of variable $X_{i_j}$ in $C_i$; and  to $0$, otherwise. For
example, for $C_6 = X_6 \lor \neg X_9 \lor X_{12}$, we store
$(C_6, X_6,1, X_9, 0,  X_{12}, 1)$. The weight of this tuple is
$2^n$.

Then the answer to the ground atomic query $\nit{Var}(X_i,1)$ is
$\nit{yes}$ iff the variable $X_i$ is assigned value $1$ in the
lexicographically maximum assignment (in case  such a satisfying
assignment exists). In case a satisfying assignment does not
exist, then the tuple in $\nit{Dummy}$ has to be changed in order
to satisfy the constraints. No attribute value in a tuple in
$\nit{Clause}$ is changed, because the cost of such a change is
higher than a change in the $\nit{Dummy}$ relation.}
}

\ignore{
\vspace{-3mm}
\defproof{Theorem \ref{prop:incr-attr-denial}}{
By reduction from the problem $P$ of CQA in \cite[theorem
4(b)]{bbfl-2005}. We introduce a new relation $\nit{Dummy}$, and
transform every denial $\forall \bar{y} \neg (A_1 \wedge \cdots
\wedge A_s)$ for  problem $P$ into $\forall \bar{y} \forall x \neg
(A_1 \wedge \cdots \wedge A_s \wedge \nit{Dummy}(x))$.
 If we start with the empty extension for $\nit{Dummy}$, the
database is consistent. On the update part, if we insert the tuple
$\nit{Dummy}(c)$ into the database, and the original denials were
inconsistent in the given instance, then we cannot delete that
tuple and no change in it can repair any violations. Thus, the
only way to repair database is  as in \cite{bbfl-2005}, which
makes CQA $P^\nit{NP}$-hard.}
}

\ignore{
\vspace{-3mm}
\defproof{Lemma \ref{lem:3col4reg}}{If a vertex $v$ in $G$ has degree 2, then
we transform it into a vertex of degree 4 by hanging from it an
``ear" as shown in the figure, which is composed of three
connected versions of the graph $H_3$ \cite[Theorem 2.3]{gjs-1976}
plus two interconnected versions of a box graph (c.f. Figure \ref{fig:fig2}).

\vspace{-2mm}
\ignore{
\begin{multicols}{2}
\vspace*{-6mm}
\psset{xunit=1mm,yunit=1mm}
\begin{pspicture}(-5,-8)(55,60) \psline(25,50)(28,58)
\psline(25,50)(22,58) 
\uput[r](25,50){$v$~~r} \psdot(20,40) \uput[r](20,40){g}
\psdot(30,40) \uput[r](30,40){b} \psdot(25,35) \uput[r](25,35){r}
\psdot(15,30) \uput[d](15,30){r} \psdot(21,30) \uput[d](21,30){b}
\psdot(29,30) \uput[d](29,30){g} \psdot(35,30) \uput[r](35,30){r}
\pscircle(15,30){5pt} \pscircle(35,30){5pt} \psline(25,50)(20,40)
\psline(25,50)(30,40) \psline(20,40)(25,35) \psline(25,35)(30,40)
\psline(20,40)(15,30) \psline(20,40)(21,30) \psline(25,35)(21,30)
\psline(25,35)(29,30) \psline(30,40)(29,30) \psline(30,40)(35,30)
\psline(15,30)(21,30) \psline(21,30)(29,30) \psline(29,30)(35,30)
\psdot(15,30) \psdot(10,20) \uput[r](10,20){g} \psdot(20,20)
\uput[r](20,20){b} \psdot(15,15) \uput[r](15,15){r} \psdot(5,10)
\uput[u](5,10){r} \psdot(11,10) \uput[u](11,10){b} \psdot(19,10)
\uput[u](19,10){g} \psdot(23,10) \uput[u](23,10){r}
\psline(15,30)(10,20) \psline(15,30)(20,20) \psline(10,20)(15,15)
\psline(15,15)(20,20) \psline(10,20)(5,10) \psline(10,20)(11,10)
\psline(15,15)(11,10) \psline(15,15)(19,10) \psline(20,20)(19,10)
\psline(20,20)(23,10) \psline(5,10)(11,10) \psline(11,10)(19,10)
\psline(19,10)(23,10) \psdot(35,30) \psdot(30,20)
\uput[r](30,20){g} \psdot(40,20) \uput[r](40,20){b} \psdot(35,15)
\uput[r](35,15){r} \psdot(27,10) \uput[u](27,10){r} \psdot(31,10)
\uput[u](31,20){b} \psdot(39,10) \uput[u](39,10){g} \psdot(45,10)
\uput[u](45,10){r} \psline(35,30)(30,20) \psline(35,30)(40,20)
\psline(30,20)(35,15) \psline(35,15)(40,20) \psline(30,20)(27,10)
\psline(30,20)(31,10) \psline(35,15)(31,10) \psline(35,15)(39,10)
\psline(40,20)(39,10) \psline(40,20)(45,10) \psline(27,10)(31,10)
\psline(31,10)(39,10) \psline(39,10)(45,10) \psdot(5,5)
\uput[d](5,5){b} \psdot(15,5) \uput[d](15,5){g} \psdot(23,5)
\uput[d](23,5){b} \psline(5,5)(15,5) \psline(15,5)(23,5)
\psline(15,5)(5,10) \psline(15,5)(23,10) \psline(5,5)(5,10)
\psline(23,5)(23,10) \psdot(27,5) \uput[d](27,5){g} \psdot(35,5)
\uput[d](35,5){b} \psdot(45,5) \uput[d](45,5){g}
\psline(27,5)(35,5) \psline(35,5)(45,5) \psline(35,5)(27,10)
\psline(35,5)(45,10) \psline(27,5)(27,10) \psline(45,5)(45,10)
\pscurve(5,5)(15,1)(27,5) \pscurve(5,5)(25,1)(45,5)
\psline(23,5)(27,5) \pscurve(23,5)(35,1)(45,5)
\end{pspicture}
}

\begin{figure}
  \centering
  \includegraphics[width=5cm]{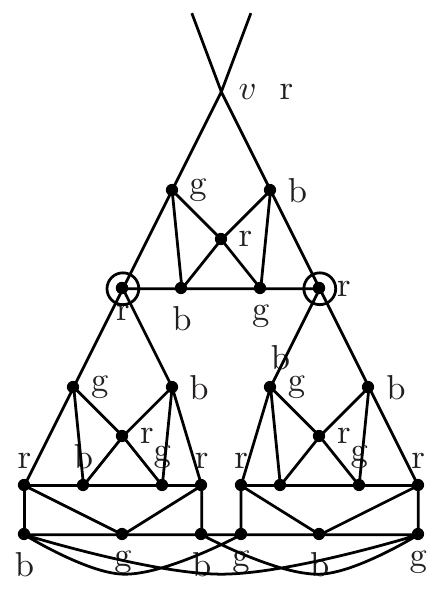}
  \caption{  }\label{fig:fig2}\vspace{-3mm}
\end{figure}

It is easy to see that the ear is regular of degree
4, is 3-colorable (as shown in Figure \ref{fig:fig2}
  with colors {\tt r,g,b}), but not
  planar. Hanging the ear adds a constant number of vertices. Now
  we have to deal with the set $V_\nit{odd}$ of vertices of degree 1 or 3 (vertices of degree 0 can be
  ignored). By Euler's theorem, $V_\nit{odd}$ has an even
  cardinality. This makes it possible to pick up
disjoint pairs $\{v_1, v_2\}$ of elements of $V_\nit{odd}$,
leaving every vertex coupled to some other vertex.
 For each such pair, $\{v_1,
v_2\}$, add an extra vertex $v'$ connected to (only) $v_1$ and
$v_2$.  This trio is 3-colorable.

Now $v_1, v_2$ have degree 2 or 4. ~~From
those that become of degree 2, hang the ``ear" as before. In this
way, all the nodes become of degree 4. The number of added
vertices is polynomial in the size of the original graph. The
4-colorability of $G'$ follows from the 4-colorability of $G$
(every planar graph is 4-colorable) and the 4-colorability of the
hanging ears.}
}

\vspace{-3mm}
  \defproof{Theorem \ref{th:incr-set-del-ins}}{If the update operation $U$ is a
  $\nit{delete}$ of
  a database atom, we reduce to our problem
  3-Colorability of
  planar graphs $G$ with vertex degree at most 4, which is
  \textit{NP}-complete \cite{gjs-1976}. Given such a non-empty graph $G$, we construct
  graph $G'$ as in
  Lemma \ref{lem:3col4reg}, which is also 4-colorable (because $G$ is and the
ears too).

Let  $E(X, Y)$ be a database relation encoding the edges of the
graph, $\nit{Coloring}$ a 2-ary database relation storing a
coloring of the vertices, and $\nit{Colors}$ a unary relation
storing the four colors
  allowed. Notice that a 4-coloring of $G$ can be found in polynomial time
  \cite{rsst-96}. Then also a 4-coloring for $G'$ can be found in polynomial
time (a 4-coloring for the ears can be given once and for all). The ICs,
essentially denials and inclusion dependencies, are as
  follows:
  \begin{enumerate}
  \item Every node is colored:~ $\forall xy \exists z (E(x,y) \rightarrow
  \nit{Coloring}(x, z))$.
  \item
  Nodes have one color: $\forall x y_1 y_2 \neg(\nit{Coloring}(x, y_1) \land
  \nit{Coloring}(x, y_2) \land y_1 \not = y_2)$.

  \item
  Colors must be allowed:~
~$\forall xy (\nit{Coloring}(x,y) \rightarrow \nit{Colors}(y))$.

  \item
  Vertex degree is not less than 4:~
  $\forall x ( \exists y E(x,y) \rightarrow \exists y_1 y_2 y_3 y_4 (E(x,y_1)
  \land E(x,
  y_2) \land E(x, y_3) \land E(x, y_4) \land y_1 \not = y_2
  \land y_1 \not = y_3 \land y_1 \not = y_4 \land y_2 \not = y_3 \land
  y_2 \not = y_3 \land y_3 \not = y_4))$.

  \item
  Vertex degree is not bigger than 5:~
  $\forall x y_1 \cdots y_5 \neg (E(x, y_1) \land \cdots \land
  E(x,y_5) \land y_1 \not = y_2 \cdots \land y_4 \not =  y_5)$.

  \item
  Only vertices are colored:~
  $\forall xy\exists z(\nit{Coloring}(x,y) \rightarrow E(x,z))$.

  \item All colors are used:~
  $\forall x\exists z (\nit{Colors}(x) \rightarrow \nit{Coloring}(z,x))$.
  \item $E$ is symmetric:~ $\forall xy(E(x,y) \rightarrow E(y,x))$.
  \item Adjacent vertices have different colors:\\ $\forall
  xyuw \neg (E(x,y) \wedge \nit{Coloring}(x,u) \wedge
  \nit{Coloring}(y,w)
\wedge u = w)$.
  \end{enumerate}
  The initial database $D$ stores the graph  $G'$, together with its
  4-coloring (that does use all 4 colors). This is a consistent instance.

  For the incremental part, if the update $U$ is the deletion of a color,
  e.g. $\nit{delete}_\nit{Colors}(c)$, i.e. of
  tuple $(c)$ from $\nit{Colors}$,  the instance becomes inconsistent, because an
inadmissible color  is being used in the coloring. Since repairs
can be obtained by changing attribute values in existing tuples
only, the only possible
  repairs are the 3-colorings of
  $G'$ with the 3 remaining colors  (if such colorings exist), which are obtained
  by changing colors in the second attribute of $\nit{Coloring}$.
  If there are no colorings, there are no repairs.

The query $Q\!:~ \nit{Colors}(c)?$ is consistently true only in
case there is no 3-coloring of the original graph $G$, because it
is true in the empty set of repairs.}

\vspace{-3mm} \defproof{Theorem \ref{theo:redCA}}{Given a schema $\cal R$
with relations $R_1, \ldots, R_m$ for CQA under the C-repair semantics,
expand each relation $R_i$ to $\bar{R}_i$ that has an extra attribute $E_i$
that takes numerical values $0$ or $1$, and is the only fixable attribute
for $\bar{R}_i$. Transform each denial of the form
$\forall \bar{x} \neg( \cdots R_i(\bar{t}) \cdots)$
into the denial
$\forall \bar{x} \cdots \forall e_i \cdots \neg(\cdots \bar{R}_i(\bar{t},e_i) \wedge e_i = 1 \cdots)$.

An atomic query $R_i(\bar{t})$ for CQA under the C-repair semantics is
transformed into $R_i(\bar{t},1)$, which is answered under the least-squares
A-repair semantics.

An instance $\bar{D}$ is created from an instance $D$ for $\cal R$, by
inserting $\bar{R}_i(\bar{c},1)$ into $\bar{D}$ when $R_i(\bar{c}) \in D$.
}

}

\end{document}